\documentclass[12pt]{elsarticle}
\usepackage{amsmath,amssymb}
\usepackage{graphics,graphicx}
\usepackage[margin=2.5cm]{geometry}
\usepackage{dcolumn,bm}
\usepackage{psfrag}
\usepackage{color}
\journal{Physica A}

\newcommand{\be}{\begin{equation}}
\newcommand{\ee}{\end{equation}}
\newcommand{\cu}
{\address{Department of Physics, University of Calcutta,
92 Acharya Prafulla Chandra Road, Kolkata 700009, India.}}

\newcommand{\imsc}
{\address{The Institute of Mathematical Sciences, CIT Campus, Taramani, Chennai 600113, India.}}

\newcommand{\epsi}{\epsilon}

\begin{document}
\begin{frontmatter}

\title{
  $A+ A \to \emptyset$ system in one 
dimension with particle motion determined by nearest neighbour 
distances: results for parallel
updates.  
}

\author{Reshmi Roy}%
\cu
\ead{reshmi.roy80@gmail.com}
\author{Parongama Sen}%
\cu
\ead{parongama@gmail.com}
\author{Purusattam Ray}%
\imsc
\ead{ray@imsc.res.in}

\begin{abstract}

A  one dimensional $A+A \to \emptyset$ system where 
the direction of motion of the particles is determined by the position of the 
nearest neighbours is studied. 
The  particles move with a probability $0.5 + \epsi$ towards their nearest neighbours with $-0.5 \leq \epsi \leq 0.5$. 
This implies  a stochastic motion towards the nearest neighbour or away from it 
for positive and negative values of $\epsi$ respectively,  with 
$\epsi = \pm ~0.5$ the two  deterministic limits.
The position of the particles are updated in parallel. 
The macroscopic as well as tagged particle dynamics are studied which show drastic changes from the 
diffusive case $\epsi=0$. 
The decay of particle density 
shows departure from the usual power law behaviour as found in $\epsi =0$, on both sides of $\epsi =0$.
The persistence probability $P(t)$ is also calculated 
that shows a power law decay, $P(t) \propto t^{-\theta}$,
for $\epsilon =0$, where $\theta \approx 0.75$, twice of what is obtained in asynchronous updating. For 
$\epsi < 0$,   $P(t)$   decays in a  stretched  exponential manner  and switches over to    a behaviour compatible with $P(t) \propto t^{-\theta} \ln t$  for   $\epsi > 0$. 
The $\epsi =0.5$ point   is characterized by the presence of permanent dimers, which are isolated pairs of particles on adjacent sites. 
Under the  parallel dynamics and for attractive interaction
these particles may go on swapping their positions for a long time, in particular, 
for $\epsi = 0.5$ these   may survive permanently.
Interestingly, for a chosen special initial condition that inhibits the formation of dimers, one recovers the asynchronous behaviour, manifesting the role of the dimers in altering the scaling behaviour for $\epsi > 0$.   
For the tagged particle, 
the  probability distribution $\Pi(x,t)$ that  the particle has undergone a displacement
$x$ at time $t$ shows the existence of a scaling variable $x/t^\nu$ where 
$\nu = 0.55 \pm 0.05$ for $\epsi > 0$ and varies with $\epsi$ for $\epsi < 0$. 
Finally,  a comparative analysis for the behaviour of all the relevant quantities for the system using  parallel and  asynchronous dynamics (studied recently) shows that there are significant differences for $\epsi > 0$ while the results are qualitatively similar  for $\epsi < 0$.

\end{abstract}

\end{frontmatter}


\section{Introduction}
\label{intro}

Reaction diffusion systems have received
a lot of attention in recent years  and have been studied in different contexts \cite{privman,krapivsky}. 
$A+A \to \emptyset$ may be the simplest example of this kind of reaction,
 where the particles $A$ diffuse and annihilate if they meet. When 
considered on a lattice, a  particle  hops to one of its neighbouring sites, and in case a particle is already there, both get annihilated. 
This reaction 
has direct mapping to the dynamical evolution of the Ising Glauber model  when
studied with asynchronous updates, i.e., when the positions of the particles
are updated one by one.

The $A+A \to \emptyset$ system has been studied in the recent past  where the particles $A$ move with a bias  
towards or away from their nearest neighbours \cite{soham_ray_ps2011, ray_ps2015,ray_daga2019,roy2020,park2020-1,park2020-2,roy_ps2020,pratik2019,park2}.  
The annihilation process is not directly affected 
by the bias which governs only the direction of motion but this extension leads to drastic changes in the  dynamical 
properties. 
The previous studies were made using asynchronous updating rule. 
 With asynchronous dynamics, the system, in a certain limit  can be mapped to an opinion dynamics problem studied
earlier \cite{soham_ps2009}. 
However, regarding the $A + A \to \emptyset$ reaction with bias as an independent problem, one can also 
consider parallel dynamics where particle positions are updated simultaneously. 
Parallel or synchronous  updating rule is an alternative way of studying 
dynamical systems and the results may vary significantly \cite{block_bergersen1999}
and hence of potential interest. 
 Essentially   time is  varied as a  discrete variable
 in the parallel update which can be relevant for social phenomena like herding behaviour 
for which such reaction diffusion systems may be regarded as a minimal model
\cite{pratik2019}.
Various physical and social phenomena  have been studied using both asynchronous and parallel dynamics and  comparative estimates show significant differences
  \cite{block_bergersen1999,Righ_takacs2014,kfir_kanter,potts,Ray_menon2001,Nareddy_machta2020}. 
One interesting exact result in the Ising Glauber and Potts models is 
 that the persistence exponent (obtained from the algebraic decay with time of 
the  probability that a spin has not  flipped till a given time \cite{bray}) is double in the case of parallel updates \cite{potts,Ray_menon2001}.

In this paper we report the results for the  dynamical 
properties of the  $A+A \to \emptyset$ system in one dimension where a  particle 
$A$ diffuses towards its 
nearest neighbour with a probability $0.5+\epsilon$ with $-0.5 \leq \epsilon\leq 0.5$ 
using parallel  dynamics.  
A similar problem  was  studied in two dimensions with parallel updates where  the 
bulk properties were considered \cite{pratik2019}. 
 Here  we study  both the macroscopic dynamical features as well as the tagged particle dynamics. 
The results, as expected,
reveal interesting differences when compared to those for asynchronous dynamics
for which both numerical  \cite{ray_ps2015,roy2020,roy_ps2020} and 
analytical results \cite{park2} are available. 
In particular, we detect a crossover in time  from the annihilation dominated regime to a regime
where the system is left  with a  constant number of particles. 

In these models, the initial condition and the type of lattice
considered are crucial factors which determine the time evolution.  The final state will depend on whether an odd number or even number of particles is present initially. The results also depend on whether the system size is odd or even when periodic boundary condition is used.  In this paper we have considered 
an even number of particles present initially as was done in the earlier 
studies, taking a lattice with even number of sites. 
Hence all the results reported here would be applicable with these conditions and we do not attempt any generalisation of the initial condition or consider  odd size lattices.
In general, 
the lattice is considered to be half-filled initially with the particles distributed randomly. However, we have also considered an  exceptional 
 initial condition where the particles occupy only the odd or even sites initially to gain a deeper insight of certain aspects of the dynamics.   

In the next section we describe the model, the dynamical scheme and simulation method. The system has very different nature for $\epsi = 0$ and positive and negative values of $\epsi$. 
For  $\epsilon > 0$ ($\epsi  < 0$), 
 the particles are biased to move towards (away from) their nearest neighbour.
The  regions $\epsi \geq 0$ and $\epsi < 0$ 
are   discussed separately in sections 3 and 4.
A comparison of the results obtained with asynchronous and parallel dynamics is presented in section 5.
Concluding remarks are made in the last section.

\section{Model and dynamical scheme,  quantities calculated  and simulation details} \label{sec_model}

The model, as mentioned in section \ref{intro}, consists of particles A 
undergoing the reaction  $A+A \to \emptyset$ in one dimension. 
We have  considered 
 lattices of size $L$   that are initially 
randomly half filled ($L$ is a multiple of 4 so that the initial number of particles is  even). 
  At each update, 
each particle hops 
one step in the direction of its nearest neighbour
with probability $0.5+\epsilon$ and in the opposite direction
with probability $0.5-\epsilon$  where $-0.5 \leq \epsilon \leq 0.5$.
If two neighbours are equidistant, 
 it moves in either direction with equal probability.
When all the particle positions are updated, one Monte Carlo (MC) step is completed.
However,   the updates are made in parallel in the sense the particle 
positions are not updated until the motion of all the particles 
have been decided, e.g., if particle X  hops from position 1 to 2, particle
Y's   motion will be decided assuming X  is at position 1 within the same 
Monte Carlo step.  
Only after  the locations of all the   particles have been updated, if   two  particles are 
found on the same lattice site,  then both of them are 
 annihilated. 

\begin{figure}[h]
\includegraphics[width= 7cm]{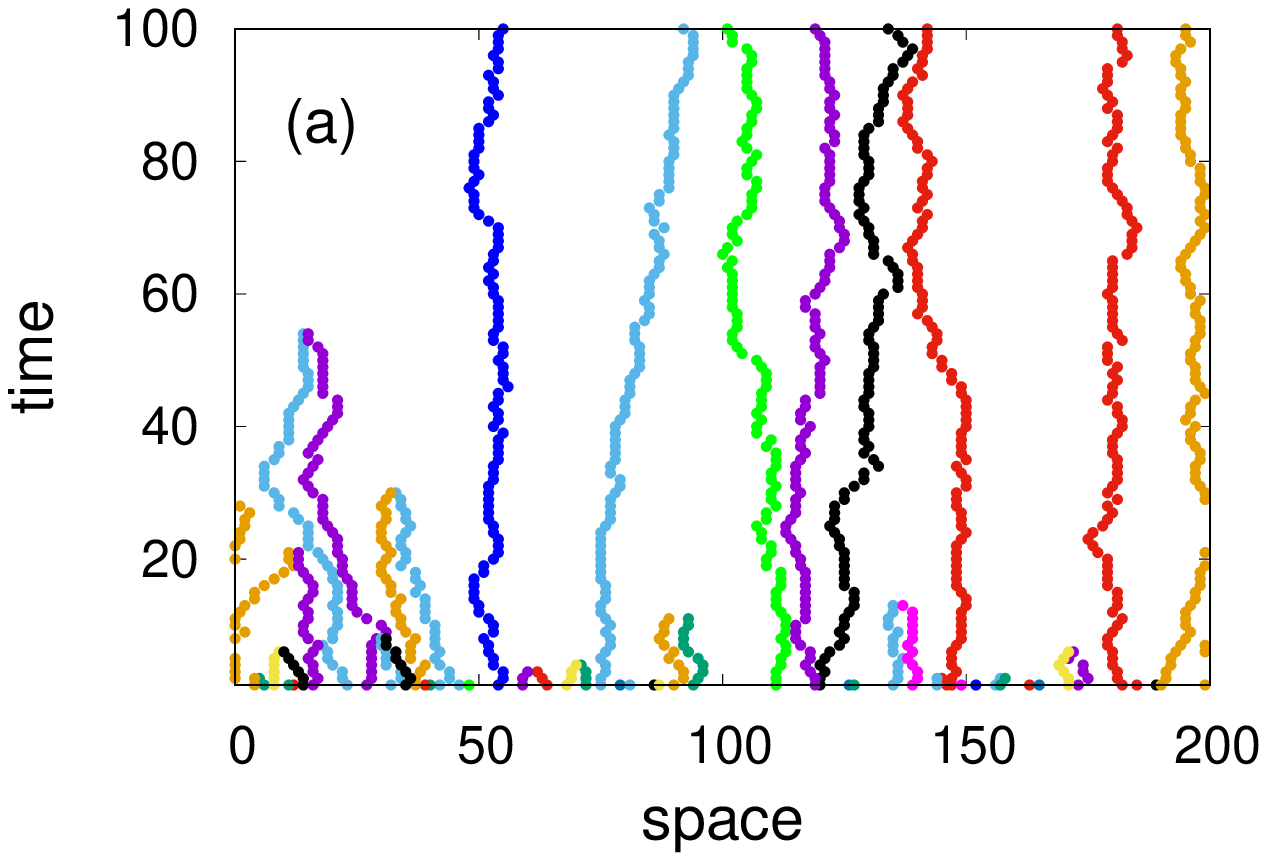} \hspace{0.5cm}
\includegraphics[width= 7cm]{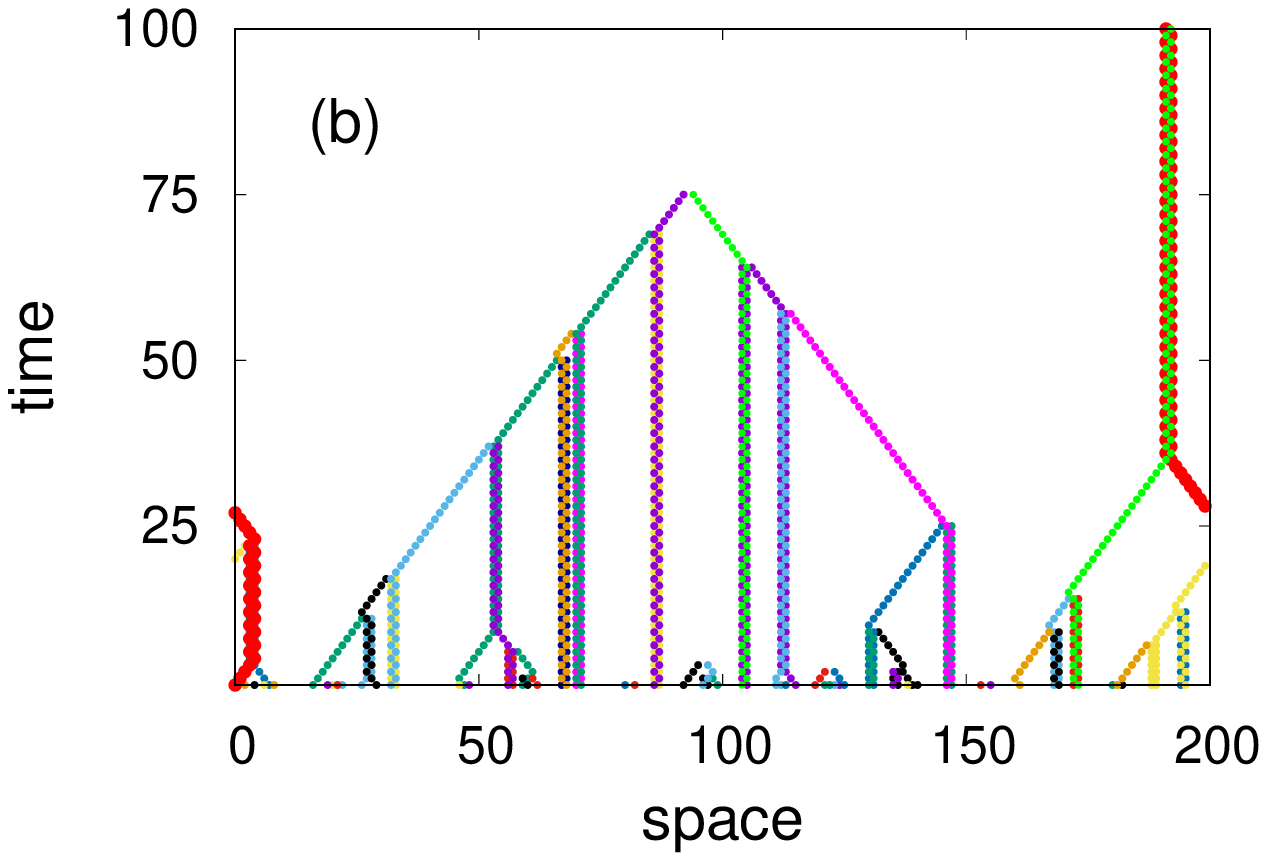}
\caption{Snapshots of the system at different times for $\epsilon=0$
(a) and $\epsilon=0.5$ (b). The
trajectories of different particles are represented by
different colors.}
\label{snappara}
\end{figure} 

For the bulk features, we have calculated the  time dependence  of the density of surviving
particles and persistence probability. 
For reaction diffusion systems, the persistence probability $P(t)$ is defined as the probability that a site has not been visited by any of the particles till time $t$.
A special feature arises for the attractive dynamics ($\epsi > 0$), namely, the formation of   dimers,  which are isolated pairs  of   particles on adjacent sites, that may go on swapping positions at every MC step for a long time.  A detailed study of such dimerisation  has been made also.

To probe the system microscopically, we have studied 
the  probability distribution $\Pi(x,t)$ that a particle has a displacement $x$ form its origin
after time $t$.  We have also estimated the  probability of change in direction of motion $S(t)$ at time $t$ and the  distribution $D(\tau)$ of time interval $\tau$ spent without change in direction of motion. 

The studies have been made  on a lattice of maximum size 32000 
and the number of realisations is generally larger for the smaller sizes; minimum number of configurations over which averaging is done is 200. In all the simulations,
periodic boundary condition has been imposed.

\section{Simulation  Results for $\epsilon \geq 0$}  
\label{results}

To get a qualitative idea of the dynamics, a plot of the   world lines of the
particles can be most helpful. Snapshots of the system are shown in Fig. 
\ref{snappara} for $\epsilon=0$ and 0.5.  It is to be noted that the motion is purely diffusive
for $\epsilon=0$ and for $\epsilon=0.5$, particles undergo  deterministic 
dynamics,
when a particle always moves towards its nearest neighbour. 
The difference in the dynamical evolution is  quite apparent; for $\epsi = 0$, one notes the usual picture of a diffusion-annihilation process 
while for $\epsi = 0.5$  two distinct behavior of the motion are manifested at long times; either the particles perform  ballistic motion  or 
pairs of  particles exist which are strongly   localised or bound. The latter is the so called dimerisation, mentioned in the last section, that happens for $\epsi > 0$. The effect of such dimerisation is maximum for $\epsi = 0.5$ where the dimers can survive for infinite times. For $\epsi < 0.5$, they may be long lived but eventually are expected to vanish.

\subsection{Bulk Properties}

\subsubsection {Fraction of surviving particles $\rho(t)$}
 
As the system evolves, the number of particles decreases due to  annihilation. 
For the purely diffusive system ($\epsilon=0$), it is well known that the fraction of surviving
particles shows a power law behaviour in time; $\rho(t)\sim t^{-\frac{1}{2}}$,
 irrespective
of the dynamics used; asynchronous or parallel. If an even number of 
particles are there initially, in the asynchronous case, at infinite times, all of them would be annihilated.  
However, in the  parallel dynamics
 there may be certain configurations where two particles will survive 
infinitely if they are separated by an odd number of lattice spacings. 
This will happen in fact for all $\epsi$ and one can expect a saturation value $\mathcal O(1/L)$ for $\rho(t)$ at $t \to \infty$. 

\begin{figure}[h]
\includegraphics[width= 8cm]{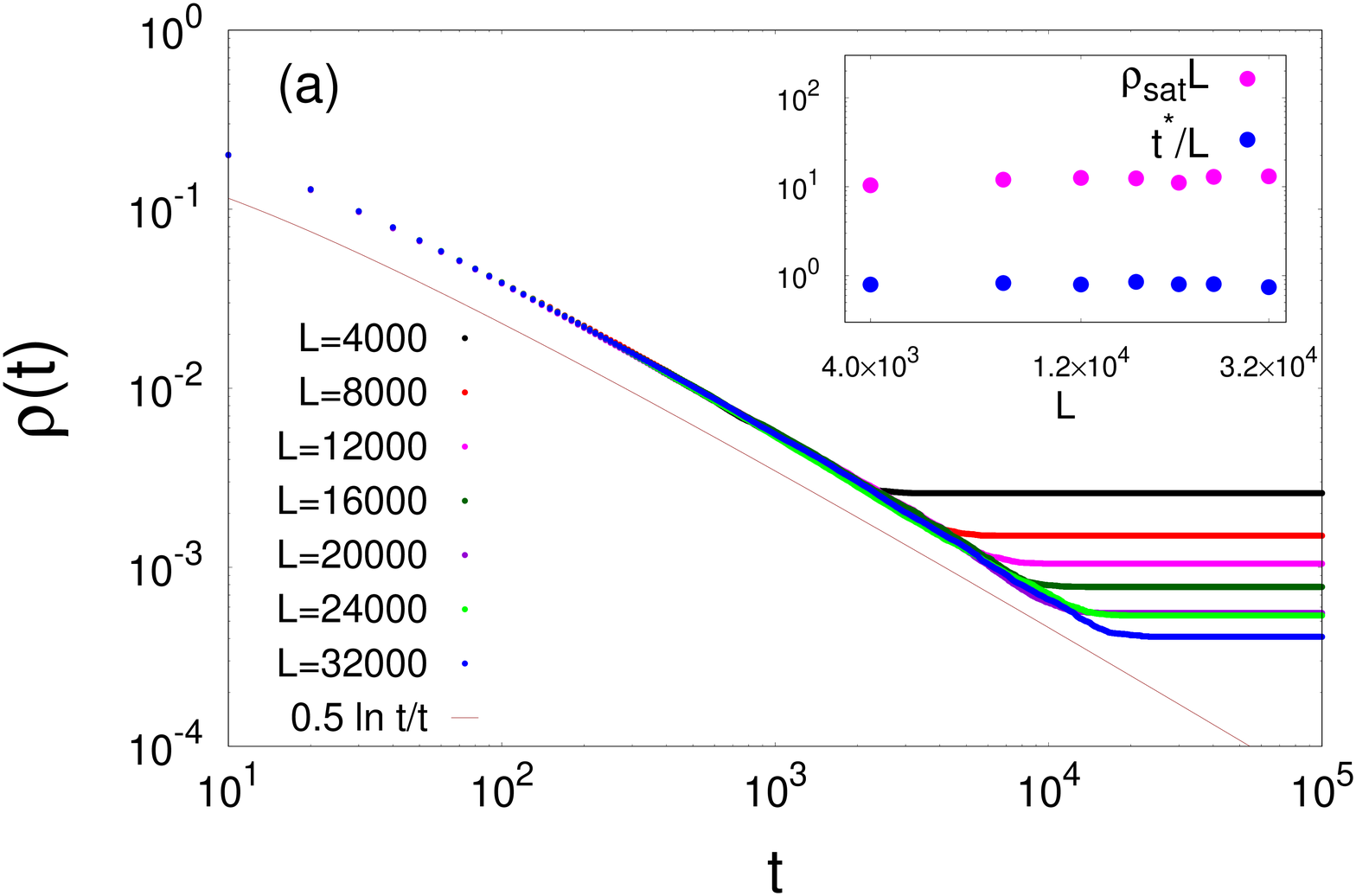} \hspace{0.7cm}
\includegraphics[width= 8cm]{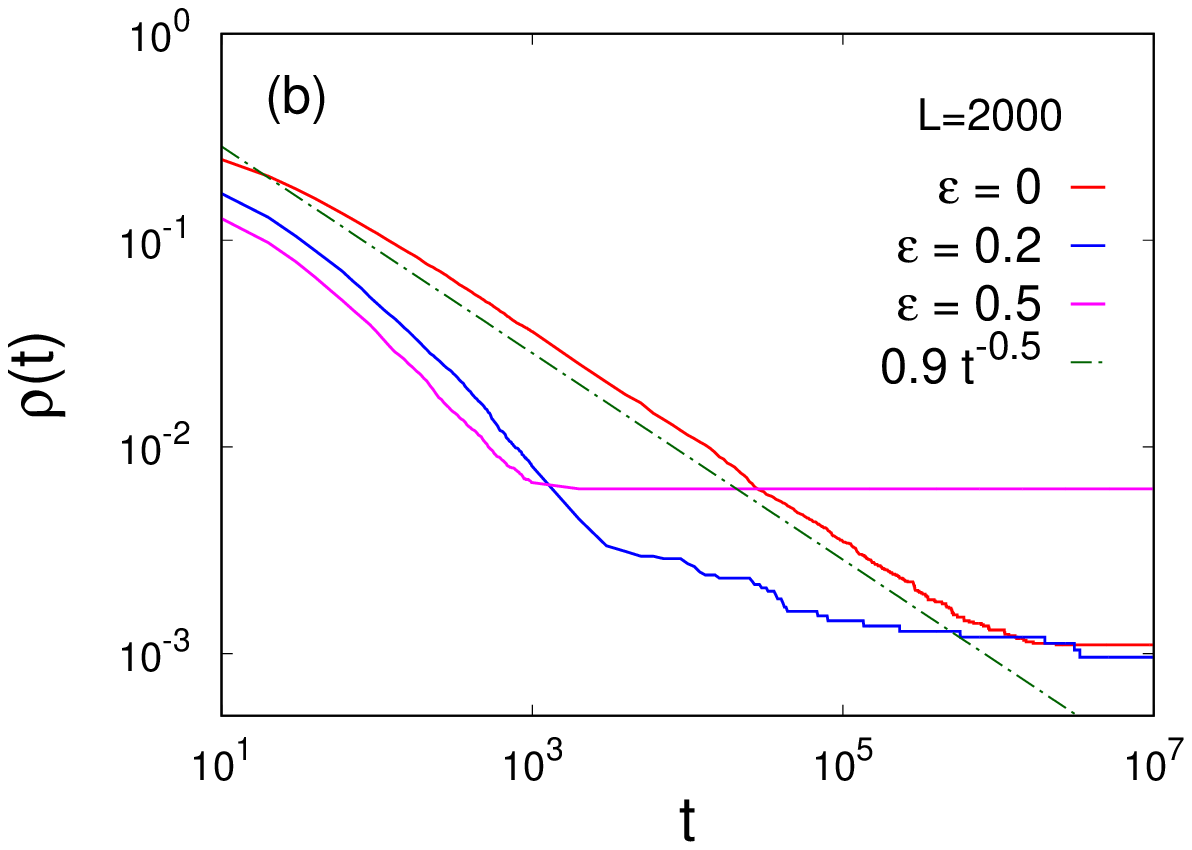}
\caption{(a) $\rho(t)$ against $t$ for $\epsilon=0.5$ is shown for several system sizes. Inset shows
the data for $\rho_{sat} L$ and 
 $t^*/L$ against system size $L$ for $\epsilon=0.5$ where $\rho_{sat}$ is the saturation value
of $\rho(t)$ and $t^*$ is the saturation time of $\rho(t)$ for $\epsilon=0.5$. The maximum number of configuration was 500.
(b) shows the data for $\rho(t)$ for a smaller system of size $L=2000$ for several $\epsilon$ simulated up to a much larger time.
The data for $\epsilon=0$ is fitted to the form $t^{-\frac{1}{2}}$.}

\label{rhopositive0.5}
\end{figure}

We discuss the case for $\epsi= 0.5$ first. 
$\rho(t)$ shows a rapid decay in time initially before abruptly 
attaining a constant value shown in Fig. \ref{rhopositive0.5}a. We have made a study for different sizes to show that the initial behaviour is independent of system size while the saturation
values $\rho_{sat} = \rho(\infty)$ are $L$ dependent. The initial decay can be fitted to a form 
\be
\rho(t) = C \ln t/t,
\label{rhot}
\ee
where $C \sim 0.85 $ independent of $L$.
The scaled saturation values $\rho_{sat} L$ has a nearly a constant  value $\mathcal O(10^1)$ for smaller $L$ values and shows a tendency to increase 
with larger $L$ shown in the inset of Fig. \ref{rhopositive0.5}a. In comparison,   $\rho_{sat}L$ for $\epsi = 0$ is $\mathcal O(1)$, shown  
for a smaller system size in Fig. \ref{rhopositive0.5}b. 
Defining  $t^*$ as the time  the saturation value is reached,  we also find that    $t^*/L$  is fairly a constant $\sim 1$ (see inset of  Fig. \ref{rhopositive0.5}a.)

\begin{figure}[h]
\includegraphics[width= 14cm]{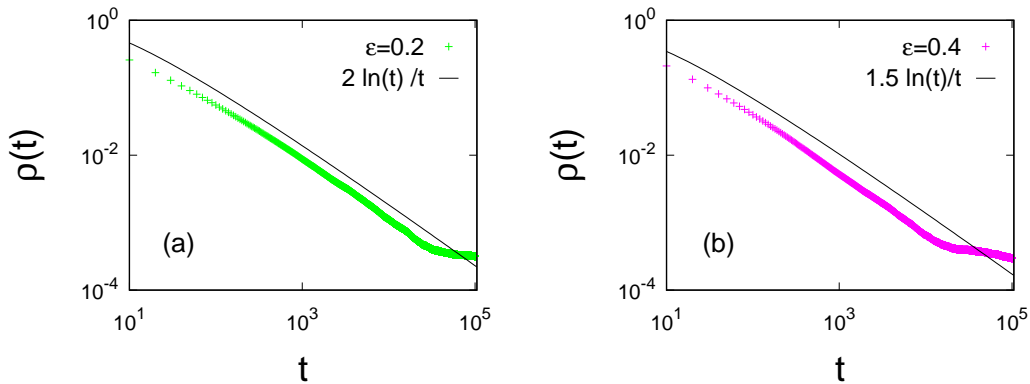}
\caption{ $\rho(t)$ versus $t$ is shown
 for several $\epsilon$ for system size $L=24000$ averaged over 200 realisations. 
 The data are fitted to
the form  $\ln t/t$. }
\label{rhopositive}
\end{figure}

For  $\epsilon \neq 0.5$,   
$\rho(t)$  can again be fitted to the form 
$\rho(t) \propto \ln t/t $. 
To check the quality of the fitting, the relative percentage error 
can be calculated as  
    $\frac{1}{T} \sum_t \frac{|\rho(t) - a \ln t/t|}{a\ln t/t} \times 100$
where $T$ is the interval of time (in the initial decay region) over which the calculation is done. 
This error   turns out to be about $5\%$ for  $\epsi = 0.2$ and $7.5\%$ for $\epsi = 0.4$ and generally of the same order for other $\epsi$ values.
$\rho(t)$ shows a slow variation beyond this initial decay region. 
Results for two values of $\epsi$ are  presented for the system size
$L = 24000 $ in Fig. \ref{rhopositive}. 
The   smaller  system size that could  be studied for a longer timescale mentioned earlier,
shows that the surviving density of particles appears to enter  a series of metastable 
regions beyond the initial faster decay and a saturation value
is  reached at very long times (Fig. \ref{rhopositive0.5}b).
 This region is difficult to reach  for larger system sizes shown in Fig. \ref{rhopositive},  
but  comparison of $\epsi = 0$ and $\epsi = 0.2$ in Fig \ref{rhopositive0.5}b shows clearly that 
$\rho(t)$ saturates at a value $\sim 1/L$ for both.

It may also be mentioned here that for the asynchronous update,  a power 
law behaviour in the  surviving fraction was  found: $\rho(t) \propto t^{-1}$ irrespective of the value of $\epsi > 0$ there. Thus the decay of the particles is 
faster when the update is made based on the current position of the particles that enhances the annihilation. 
Indeed, the $\ln t/t$  variation is rather unconventional for reaction diffusion systems.  For
$\epsi = 0.5$, the particles perform more or less ballistic motion except for the cases 
when two adjacent particles get entangled to form a dimer and continuously swap their positions. So apparently the   power law decay 
is slowed down manifested by the presence of the $\ln t$ term entering as a multiplicative factor and that can  possibly be due to the dimers, 
which do not  move ballistically and can be long lived. 
For $ 0 < \epsi < 0.5$, the dimers, though not permanent, can similarly  slow down the decay of $\rho(t)$. Detailed discussion on the dimers appears later in the paper, in Section \ref{secdimer}.

\begin{figure}[h]
\includegraphics[width= 7.5cm]{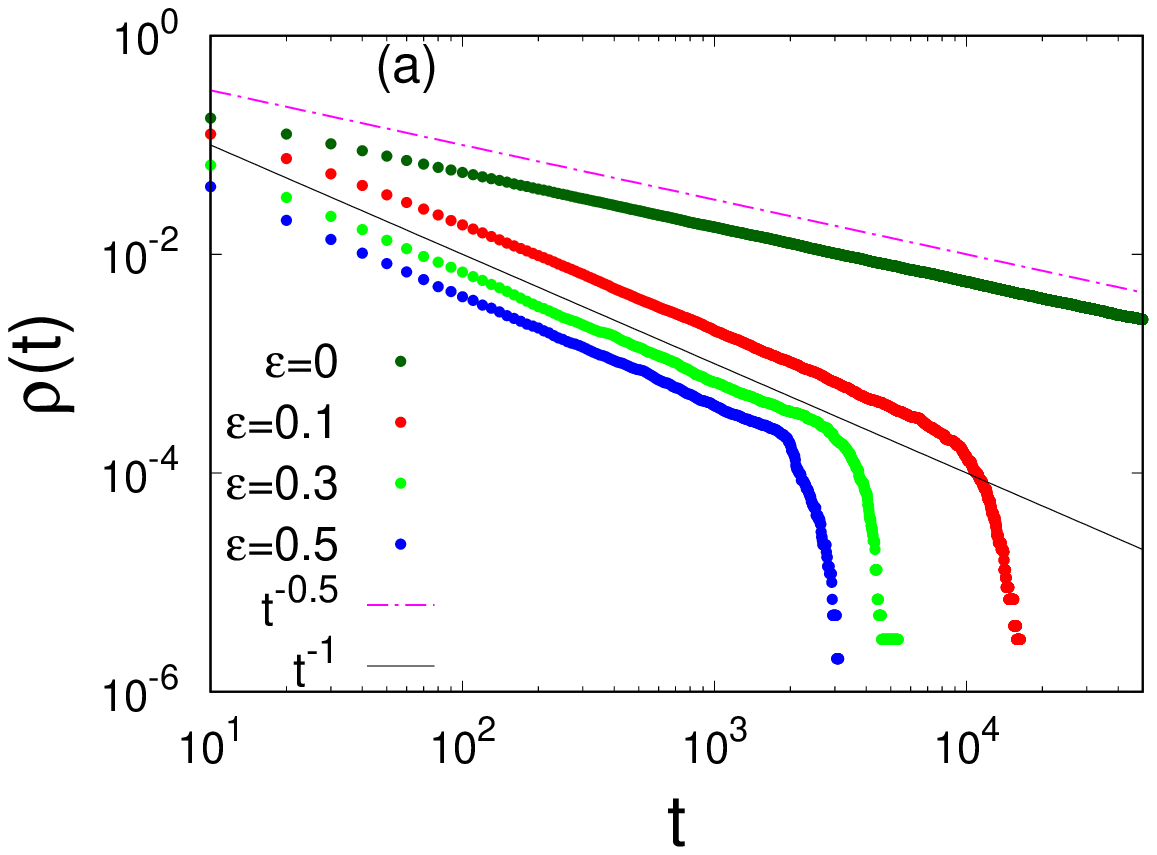}\hspace{-3.0cm}
\includegraphics[width= 7.5cm]{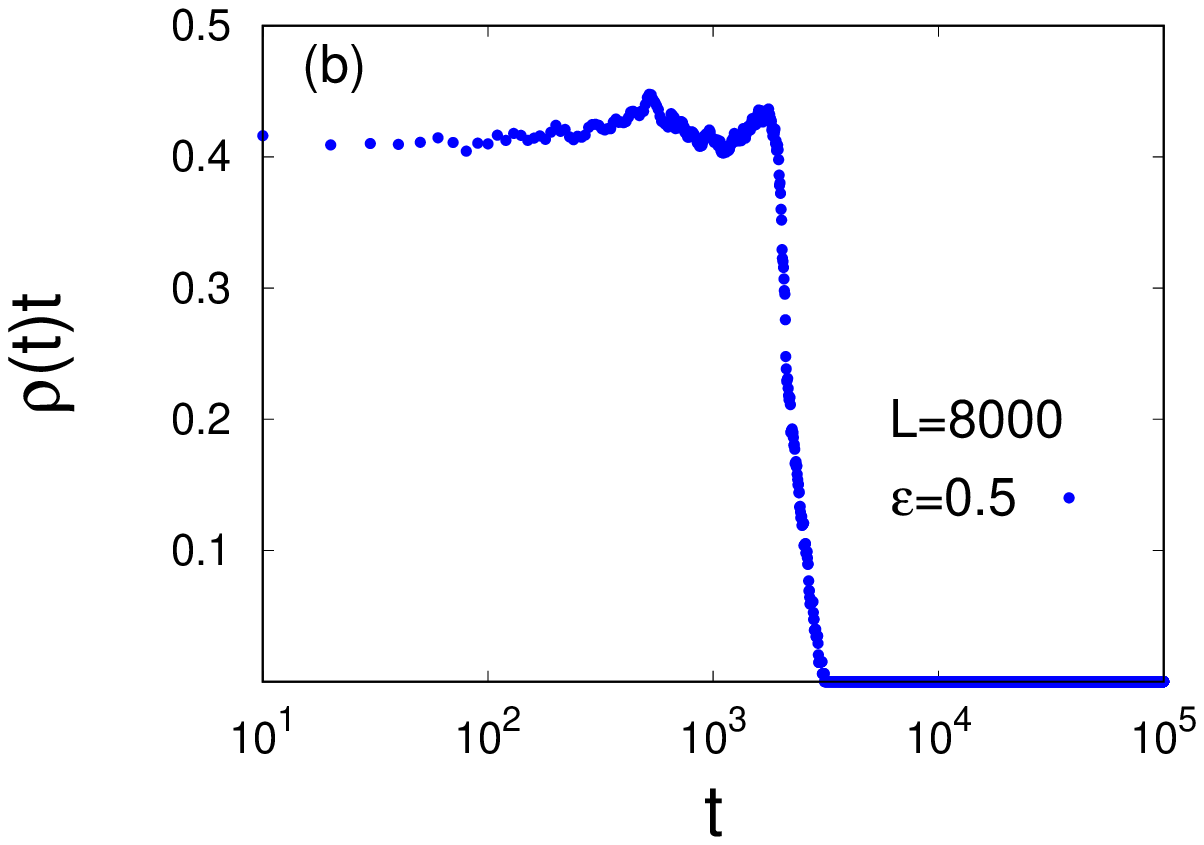}
\includegraphics[width= 7.5cm]{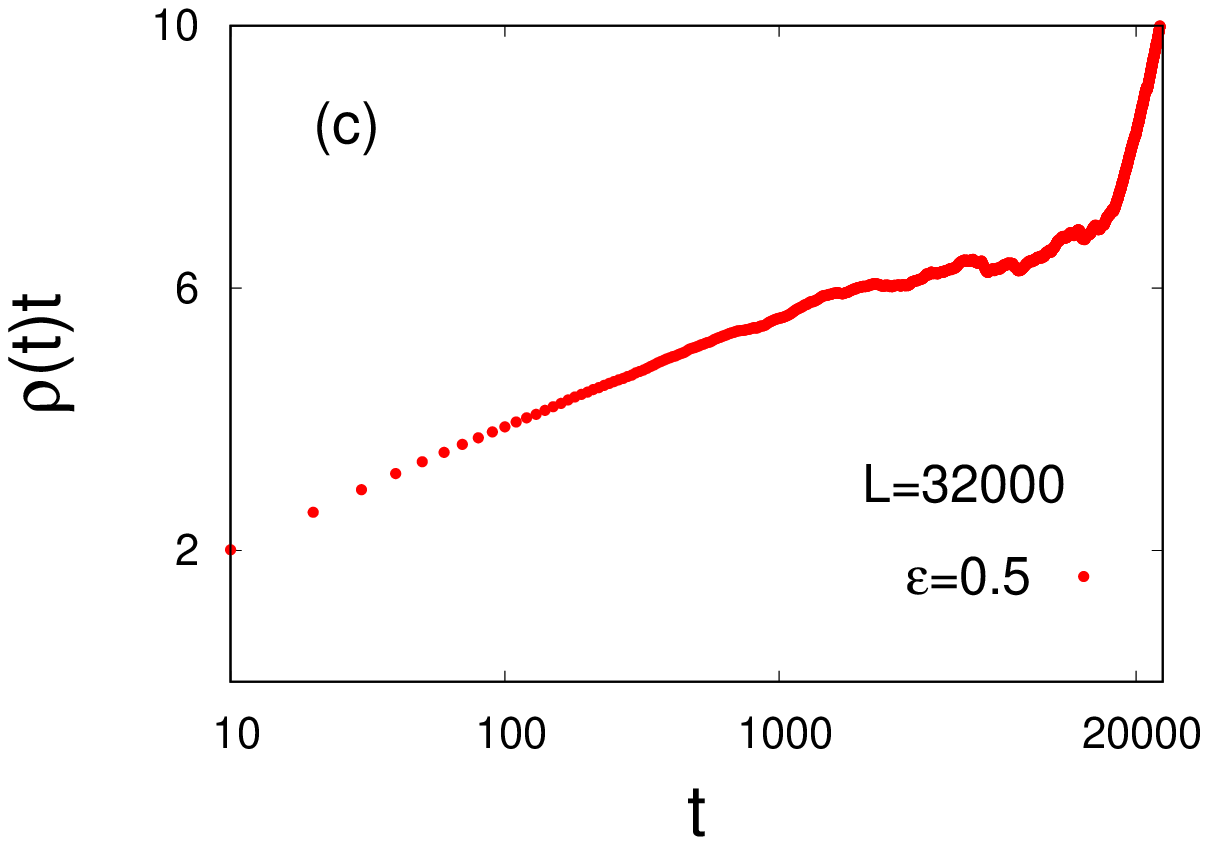}\hspace{1.3cm}
\includegraphics[width= 7.5cm]{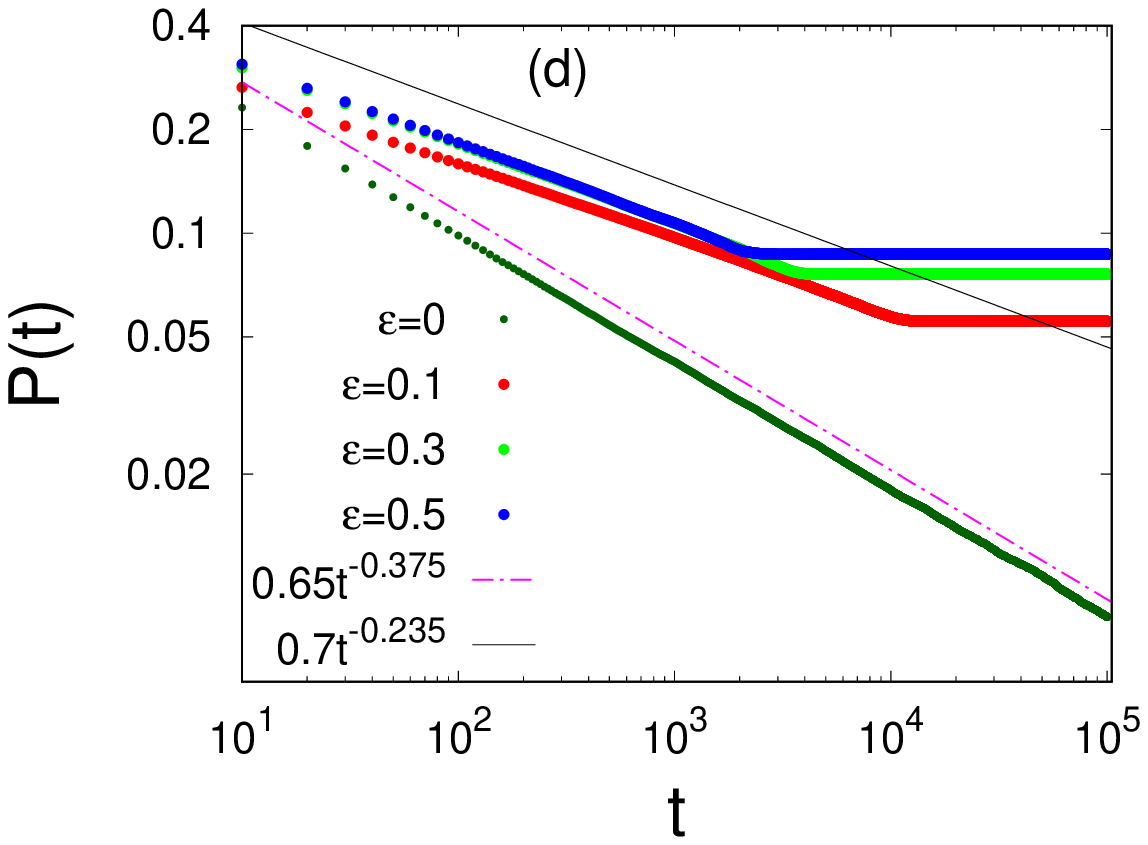}
\caption{
Results  for a system size $L=8000$ are shown in (a), (b), (d) where only
the even sites are occupied initially, averaged over 200 realisations. For (c), lattice was randomly half filled initially.
(a) Variation of $\rho(t)$ is shown against $t$
for $\epsilon=0,0.1,0.3$ and 0.5 which show power law decay with exponents $0.5$ for $\epsilon=0$ and
and $1$ for $\epsilon>0$ which is similar to the asynchronous updating results where the lattice was randomly half filled with particles.
(b) shows the semi-logarithmic plot of $\rho(t)t$ vs $t$ for $\epsilon=0.5$.
(c) Semi-logarithmic plot of $\rho(t)$ vs $t$ is shown for $\epsilon=0.5$ when the lattice was randomly half filled initially. It clearly shows a linear behavior
before it reaches the saturation region. (d) $P(t)$ plotted against $t$ for several $\epsilon$  show 
 power law decay with exponent $0.375$ for $\epsilon=0$ and $0.235$ for $\epsilon>0$, the same exponents were obtained for
asynchronous dynamics also.}
\label{alive_per_even}
\end{figure}

To eliminate the effect of dimers, we introduced a biased initial condition where all the particles are either on odd  sites or even sites. Here  it is obvious that dimer formation cannot take place and 
  one gets a nice agreement with a power law decay as $t^{-1}$ for $\epsilon>0$ shown in Fig. \ref{alive_per_even}a, confirming that the dimer formation is responsible for the deviation from a simple power law for the random 
case. For $\epsilon=0$,  $\rho(t) \sim t^{-0.5}$ is still valid. This initial condition effectively makes the system equivalent to the one with asynchronous
dynamics. Semi-logarithmic plot of $\rho(t)t$ against $t$ in Fig. \ref{alive_per_even}(b) and (c) clearly show the differences in the behaviour of $\rho(t)$ for these two cases with different initial conditions.

In this context, one may mention that   the    
decay kinetics of ballistic annihilating particles and its  several   variants 
 show power law behaviour \cite{Ben-Naim,Ben-redner,Biswas-leyvraz}  with exponents in general $\leq 1$.
We also conjecture that the early time behaviour is annihilation dominated while the later time behaviour is 
due to the presence of dimers which makes the saturation value higher than $\mathcal O(1/L)$ for $\epsi = 0.5$, which is normally expected in the system.

\subsubsection{Persistence probability $P(t)$}

Persistence probability $P(t)$, as already mentioned,  is defined as the probability that a site is unvisited till time $t$ by any of the particles.
Initially, all sites are regarded as unvisited even if a particle is put there.
When the system is updated using asynchronous dynamics,  
$P(t)$ shows a power law decay with time for $\epsilon \geq 0$,
$P(t) \sim t^{-\theta}$ with $\theta=0.375$ for $\epsilon=0$ (exact result) and $\theta \approx 0.235$ for $\epsilon>0$ \cite {soham_ray_ps2011}.
For the system with parallel updating without bias ($\epsi = 0$), we find that $\theta$ is $\sim 0.75$, a value twice of the  one obtained with  asynchronous updates. 
Such a relation of the exponent values for 
asynchronous and parallel update could be  established for the Potts and Ising models in \cite{potts,Ray_menon2001}, however, for the reaction diffusion model 
it is not obvious. 

\begin{figure}[h]
\centering
\includegraphics[width= 9cm]{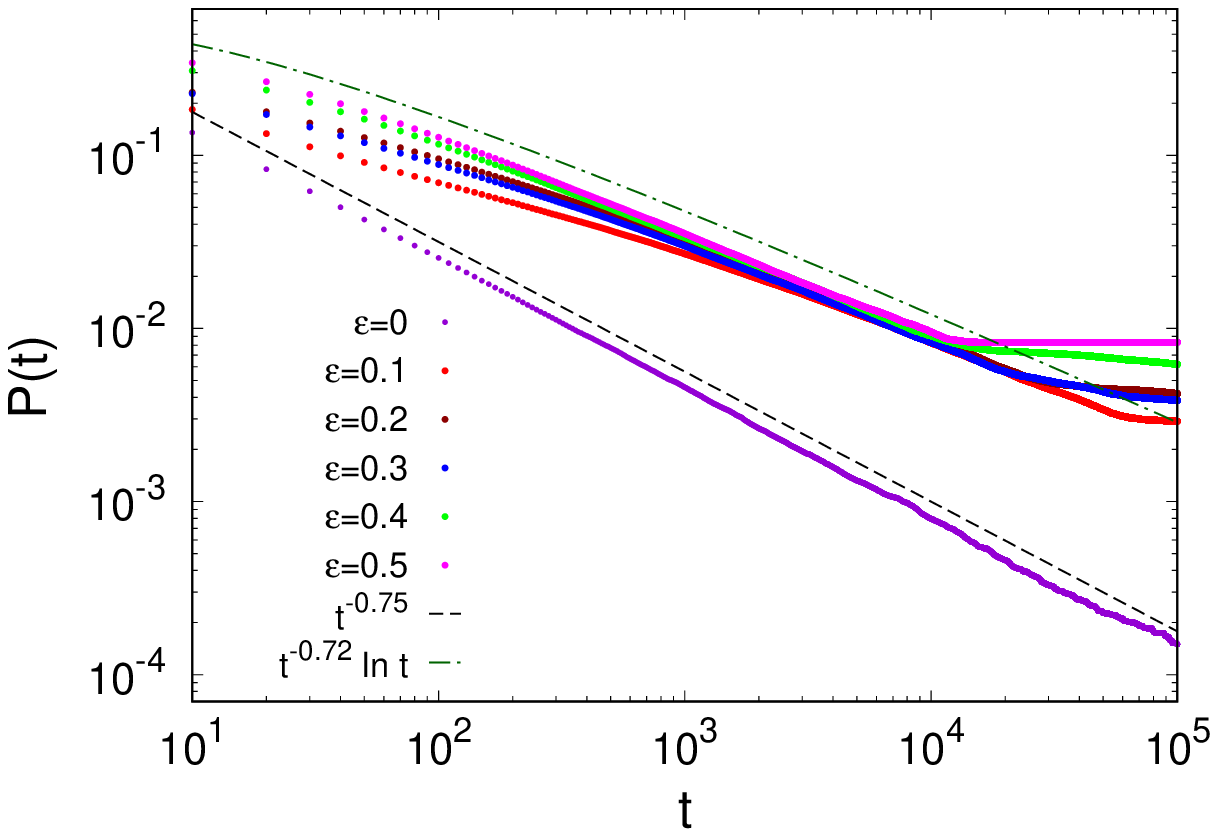}
\caption{Variation of persistence probability $P(t)$ with $t$ for different $\epsilon$ for a system size $L=24000$  averaged over 200 configurations and
the data are fitted to the form of Eq. \ref{persiseq} for $\epsi \neq 0$. For $\epsi =0$, a simple power law decay exists.}
\label{persis}
\end{figure}

For $\epsilon>0$,  $P(t)$ does not show a clear power law decay, 
a fitting of the form
\be 
P(t) \propto t^{-\theta} \ln t
\label{persiseq}
\ee
with $\theta \approx 0.72 $ seems appropriate here, shown in Fig. \ref{persis}. Hence, it appears that the 
$\epsi =0$ behaviour of $P(t)$,  with parallel updates, is modified by a factor of $\ln t$ for $\epsi \neq 0$.

We also estimate the persistence probability  with the special initial condition
with  only even (odd) sites 
occupied initially with  parallel dynamics. 
The persistence probability here shows the behaviour $P(t) \propto t^{-\theta}$ with $\theta = 0.375$
for $\epsilon=0$ and
$\theta =  0.235$ for $\epsi > 0$ with  a high degree of accuracy, the same values obtained for the 
asynchronous case.  The results are shown in Fig. \ref{alive_per_even}b. We will come back to 
this point later in the last section.

\subsubsection{Dimer Formation }
\label{secdimer}

 A dimer is an isolated pair of particles occupying two adjacent  sites, having no other neighbouring particles.
Let us consider the case for $\epsi = 0.5$ with two particles at positions $x_1$ and $x_2 = x_1+1$. Then the particle at position $x_1$ ($x_2$) will 
shift towards its nearest neighbour's   position, that is,  $x_2$ ($x_1$) due to the 
attractive force and the  particles  will go on swapping their  positions 
 unless at least one of them is annihilated by a third particle 
coming in the vicinity of either of them.
As two particles separated by an odd number of lattice spacing remaining
in the system will never be annihilated, one or more dimers are expected 
to exist  forever with a finite probability for $\epsi = 0.5$.  These dimers 
will not interact with each other and if the particles are indistinguishable,
the system would appear to reach an absorbing state. 
 We get evidence that such dimers do remain in the system from the $\rho(t \to \infty) $ data as it reaches a value of the order of $10/L$ (or higher as  $L$ increases) rather than $1/L$ to be expected in a finite system. 
 For any other value of $\epsi$,  such dimers can form but there is always a probability, however small,  that the constituent particles   
move apart, such that the system may  remain in an active state.  The  probability of dimerisation at infinite time is expected  vanish for  $\epsilon \neq 0$.
This is  consistent with the fact that $\rho(t)$ for $\epsi = 0.2$ and 0 reach the same saturation value (Fig. \ref{rhopositive0.5}b).
For $\epsi$ close to 0.5, one can expect dimers to remain at large times, however, they are not `permanent' as in the case of $\epsi = 0.5$.  

We have studied the dimer density $\langle {\rho_d}(t)\rangle $, defined as the average number of dimers 
 divided by the system size
 for several $\epsilon$.  
We note that indeed for $\epsi=0.5$,  $\langle {\rho_d}(t)\rangle $ reaches a saturation.  
As already mentioned, the dynamics become extremely slow for $\epsi < 0.5$, the data show metastable regions, however, since there is a diffusive component, 
it is expected that  dimers will not survive for infinite times. Up to the 
time studied in the simulation, the data for $\langle \rho_d(t) \rangle$ for $\epsi < 0.5$ 
indeed show a  tendency to decrease, albeit very slowly.
 
\begin{figure}[h]
\includegraphics[width=8cm]{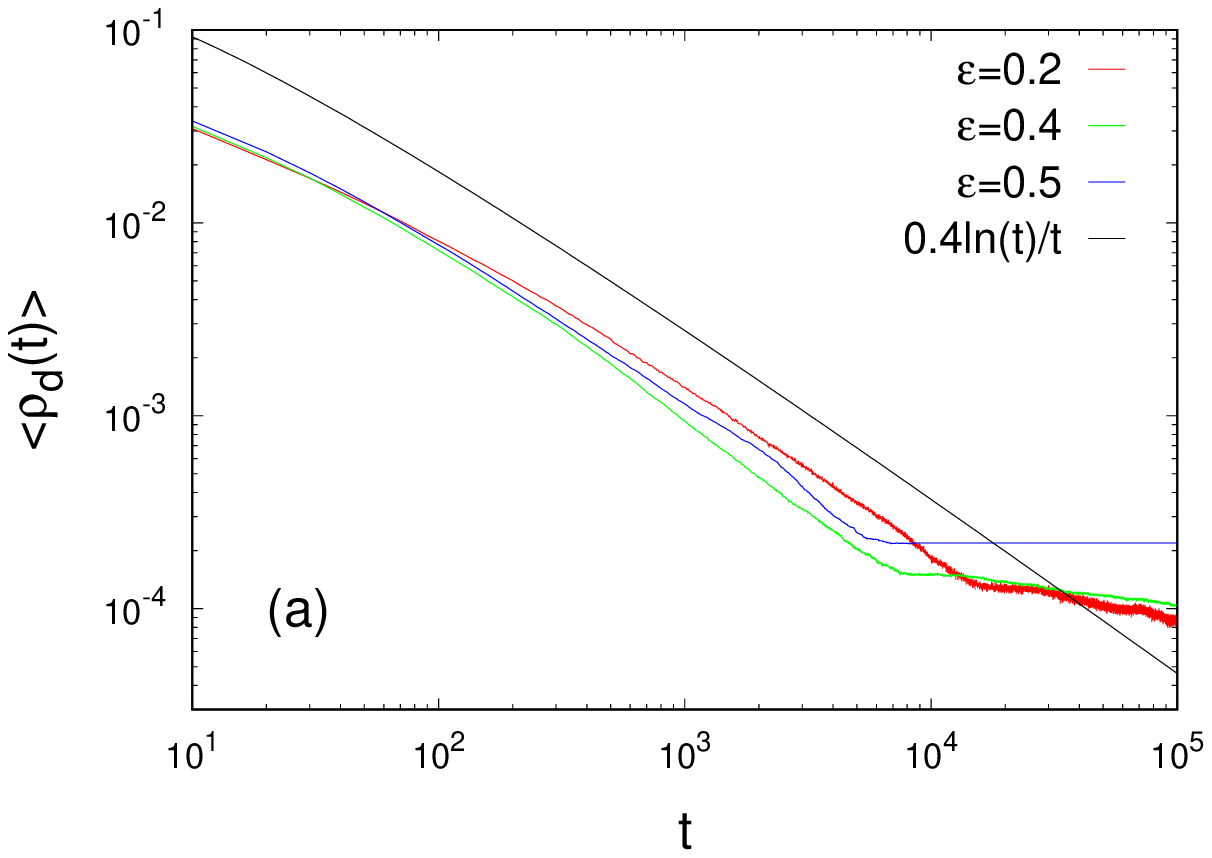}
\includegraphics[width=8cm]{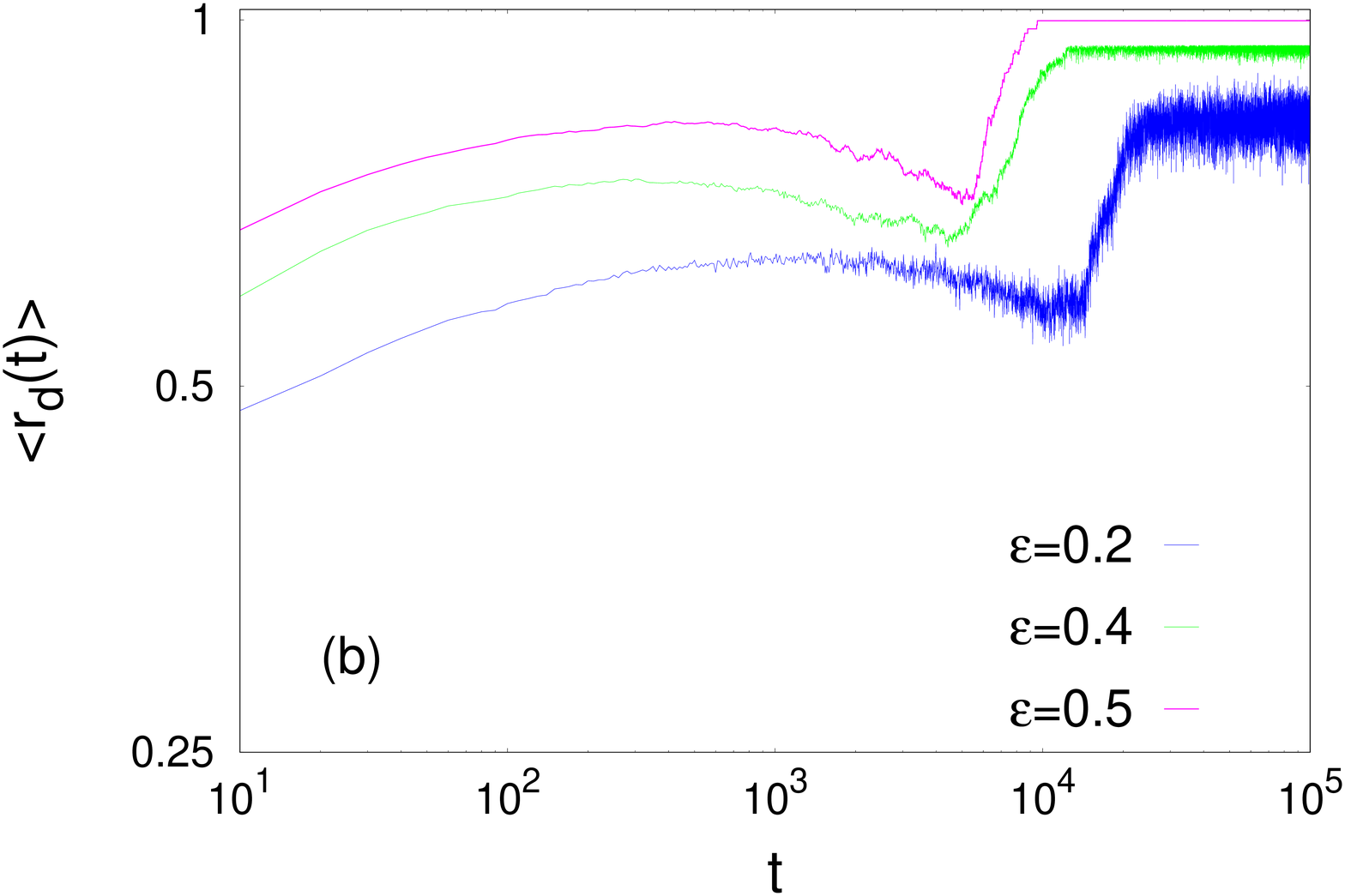}
\caption{(a) shows the variation of $\langle \rho_d(t) \rangle $ with time for several $\epsilon$. (b) shows
the variation of $\langle r_d(t) \rangle$ as a function of time for several $\epsilon$.
These data are for system size $L=12000$ averaged over 200 configurations. }
\label{dimer_time}
\end{figure}

We also find that 
an approximate fitting can be made; $\langle {\rho_d}(t)\rangle$
decays with a behaviour very close to   $ \ln t/t$ for all  $\epsi$  before it reaches a saturation region or enters the metastable regime.  
 Hence, even if dimers are not permanent for $\epsi < 0.5$, the fact that  the decay of both $\rho(t)$ 
and $P(t)$ are made slower by a factor of   $\ln t $ for all $\epsi \neq 0$, 
suggests that there is an effect  of the   dimer formation up to a long time. The data for $\rho_d(t)$ is shown in Fig. \ref{dimer_time}a.

We  study another   quantity $\langle r_d(t)\rangle$ to estimate the probability of
forming a dimer out of the 
 surviving particles at any time. $r_d(t)$ for a particular configuration is 
a ratio defined as 
\begin{eqnarray}
r_d(t)=\frac {2L \rho_d (t)}{N(t)},
\label{dimereq}
\end{eqnarray}
where $N(t)$ is the number of surviving particles 
at time $t$, calculated for $N(t) \neq 0$. 
Note that if all the particles form dimers, $\rho_d (t) = N(t)/2L$ and $r_d(t) = 1$. To calculate the 
average $\langle r_d(t) \rangle$, we take only those configurations for which $N(t) \neq 0$ at time $t$. Fig. \ref{dimer_time}b shows that $\langle r_d(t)\rangle$ has a non monotonic behaviour,  initially it increases with time and then decreases slowly. The decrease continues till it 
 shows a  dip beyond which it increases  rather sharply and ultimately   saturates  
at long times.  We conjecture that the dip occurs at around 
the time, when the particle density
reaches the phase of either very slow decay (for $\epsi < 0.5$) or becomes a constant (for $\epsi = 0.5$). 
We note here that $\langle r_d(t)\rangle$ reaches the value unity for $\epsi = 0.5$ at large times indicating all the surviving particles form dimer. For $\epsi < 0.5$ but close to it, there is a fluctuation
about a value close to 1, indicating that the dimers are not permanent as they  
form and break away regularly keeping a fairly constant value in time.  
Fig. \ref{dimer_time}b shows  the results. 

\subsection{Tagged particle features}

\subsubsection {Probability distribution $\Pi(x, t)$}

For pure random walk ($\epsilon=0$), the probability distribution $\Pi(x,t)$ 
that a particle,starting from the origin, is at position $x$ at time $t$  
is known to be Gaussian, i.e., $\Pi(x,t) \propto \frac{1}{\sqrt{t}} e^{-\frac{x^2/t}{2\sigma^2}}$. Consequently, $\Pi(x,t)t^{1/2}$ 
shows a data collapse for
different times when plotted against $x/t^{1/2}$. 
This is also true for the
unbiased annihilating random  walkers
 because they perform purely diffusive motion until they are annihilated. 

For the reaction diffusion models, one can tag a particle and trace its motion to find the displacement $x$ from its origin at any time $t$.  To obtain the 
distribution $\Pi(x,t)$, the fraction of the  surviving particles 
that underwent the same displacement $x$ at time $t$ is estimated.
 
\begin{figure}[h]
\centering
\includegraphics[width=12cm]{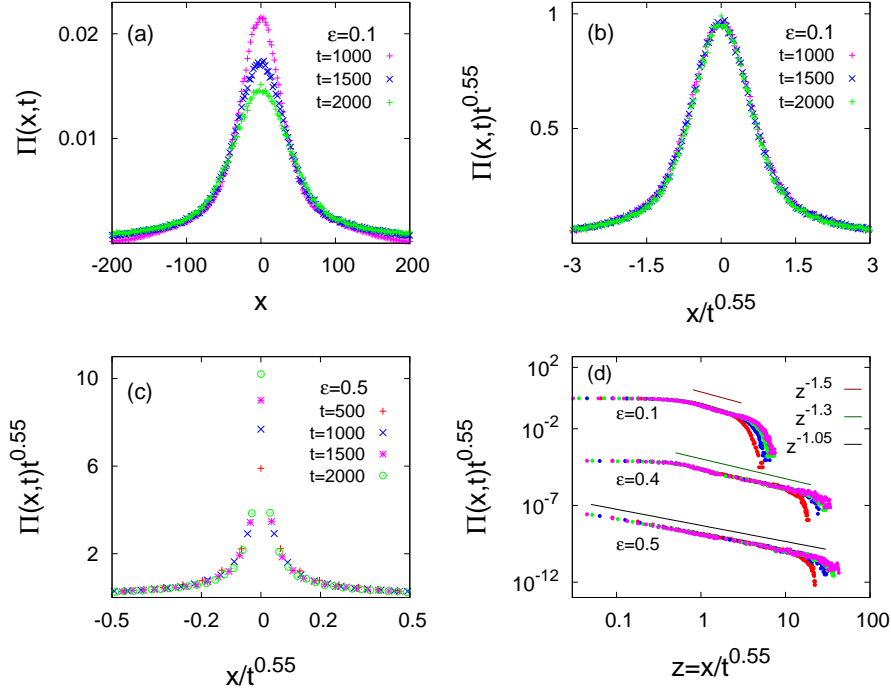}
\caption{(a) Raw data for $\Pi(x,t)$ for $\epsi = 0.1$. (b), (c) Data collapse of $\Pi(x,t)t^{0.55}$ against $x/t^{0.55}$ for $\epsilon=0.1, 0.5$ for different
times $t$. (d) Data collapse of $\Pi(x,t)$ in log-log plot reveals a power law region.  Data shown for the different $\epsi$ values  are shifted along $Y$ axis for clarity.
These results are for system size $L=12000$ averaged over 6000 different realisations.}
\label{distpara}
\end{figure}

For $\epsilon>0$,
 the distributions show a non-Gaussian single peaked structure. However, a data 
collapse can be obtained by plotting $\Pi(x,t)t^\nu$ against $x/t^\nu$ where $\nu=0.55 \pm 0.05$.
Figs \ref{distpara}a, b, c  show the raw data for $\Pi(x,t)$ against $t$  (for $\epsi = 0.1$) as well as the 
 collapsed data 
$\Pi(x,t)t^{0.55}$ against $x/t^{0.55}$ 
for $\epsi = 0.1$ and $0.5$ in a linear plot.   Fig. \ref{distpara}d shows the 
collapsed data for different values of $\epsilon$ in a log-log plot. It reveals that 
the scaling function has a constant part for small values of its argument, then it enters a power law region before reaching  a cutoff value. The cutoff 
value  increases with $\epsilon$ and also with time for each $\epsilon$.  The constant part shrinks as $\epsi$ increases (it is almost nonexistent for $\epsi = 0.5$) while the power law regime increases. The associated exponent value, mentioned in Fig.  \ref{distpara}d,  decreases with $\epsi$. The significance of these  features will be discussed in detail in sec. \ref{comparison} after the  
results  for all the other tagged particle dynamics are reported.

\subsubsection {Probability of direction change $S(t)$}

The probability of direction change $S(t)$ at time $t$ is calculated  by studying
the number of particles that change their direction of motion at time $t$ scaled  
by the total number of surviving particles at that instant of time. For pure random walk 
($\epsilon=0$) $S(t)=0.5$, independent of time. 

\begin{figure}[h]
\centering
\includegraphics[width= 9cm]{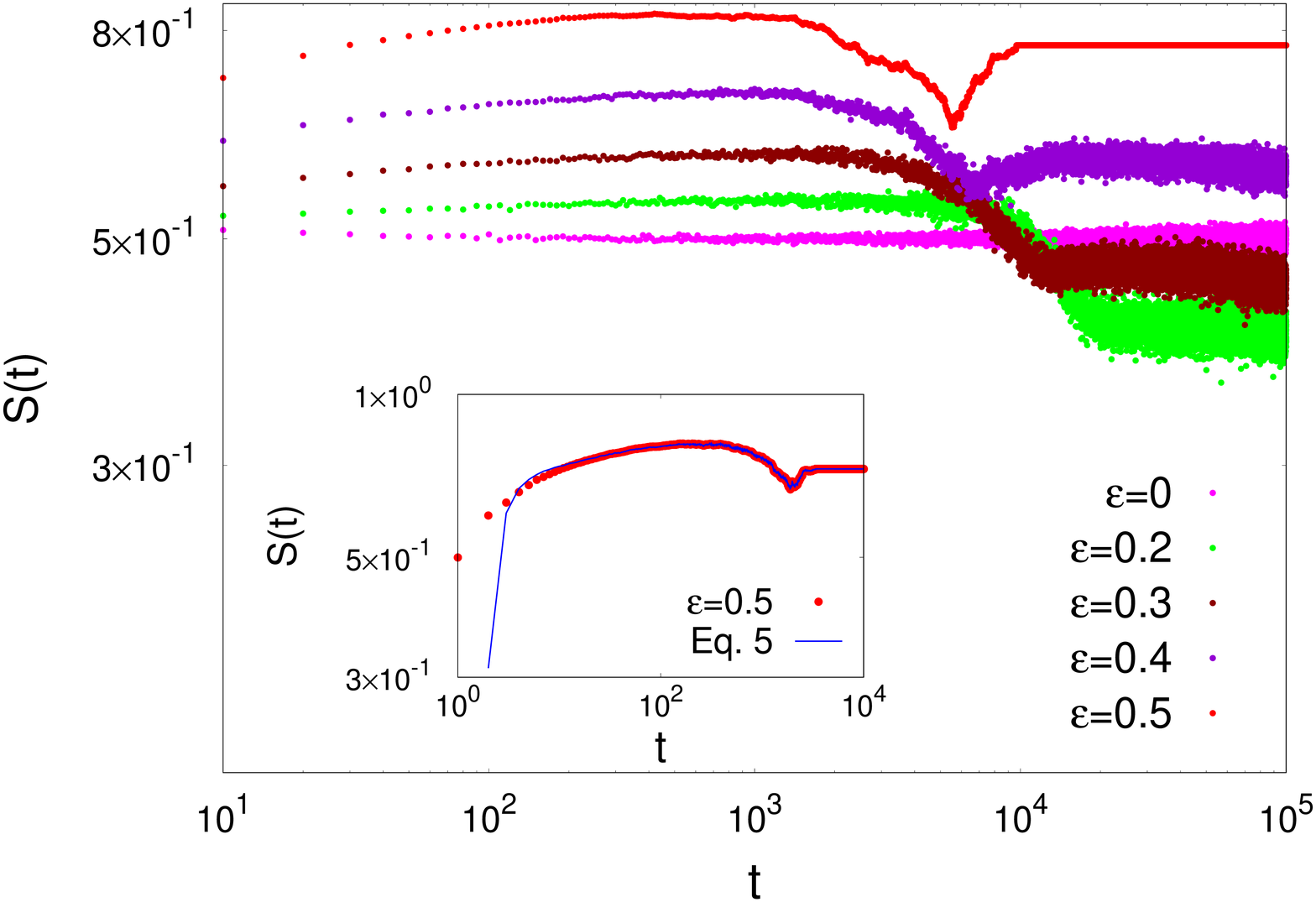}
\caption{Probability of direction change $S(t)$ of tagged particles at time $t$ for
several $\epsilon$ for a system size $L=12000$ averaged over 500 initial configurations.
 Inset shows $S(t)$ data for $\epsi = 0.5$ compared with the proposed 
form given in Eq. \ref{S(t)-t}. Number of realisation studied was 500 for this data with $L = 4000$.}
\label{dirchangepara}
\end{figure}

$S(t)$ shows an increase till a certain time  and then starts  decreasing. 
For larger $\epsi$ values, it is possible to detect a 
dip occurring subsequently, 
  beyond which $S(t)$  increases  again and attains a constant value. 
The results are shown in Fig. \ref{dirchangepara}.

We try to obtain an analytical form of $S(t)$ 
for $\epsilon=0.5$, where a particle  can change its direction of motion due to two reasons: if its nearest neighbour is annihilated (though it is a necessary
 but not sufficient condition) and/or due to dimer formation. 
Let us first estimate  the contribution to $S(t)$ due to   annihilation. 
  The change in direction due to annihilation 
per particle  is proportional to $\frac {A(t)}{\rho(t)}$ where $A(t)$ is
the number of annihilation $(A(t) \propto -\frac {d\rho}{dt})$. Therefore, we get a contribution $\propto \frac{d\rho(t)/dt}{\rho(t)}$.
The form of $\rho (t)$ is taken from Eq. \ref{rhot}. 
 Thus the  contribution from the annihilation process 
to $S(t)$ written as $S(t)_{ann}$ is given by 
\begin{equation}
S(t)_{ann} \propto  \frac{1}{t} - \frac{1}{t \ln t}. 
\label{Stann}
\end{equation}

In addition to this, contribution from the dimers should be taken into account. 
The particles 
forming a dimer will necessarily change direction at every step. 
The contribution
to $S(t)$ due to dimers will therefore be simply $r_d(t)$, the latter being the probability that there is a dimer 
and since  it will come from those configurations only which have  surviving particles 
till that time,  we need a multiplicative factor. This is because $S(t)$ is a quantity averaged over all configurations. 


Considering both contributions, 
\begin{eqnarray}
S(t) =  c_1\bigg[\frac{1}{t} - \frac{1}{t \ln t} \bigg] + c_2(t) \langle r_d(t)\rangle.
\label{S(t)-t}
\end{eqnarray}
Here $c_1$ is a proportionality constant 
 and $c_2(t)$ denotes the fraction of configurations which have $\rho(t) \neq 0$.  
%

We plot the rhs of Eq. \ref{S(t)-t} using the data for $c_2(t)$ and $\langle r_d(t)\rangle $ from the simulation and with $c_1 =1$ we get
very good agreement with $S(t)$ obtained from the simulation beyond a very short initial time (inset of Fig. \ref{dirchangepara}). 
A comparison with $\langle r_d(t)\rangle$ (Fig. \ref{dimer_time}b) also reveals the fact that $S(t)$ is  annihilation 
dominated initially but  crosses over to a regime dominated by   the  ``dimerised'' absorbing states for $\epsi= 0.5$.

\subsubsection{Distribution of time interval spent without change in direction of motion $D(\tau)$}

Another interesting quantity is $D(\tau)$, the interval of time $\tau$
spent without change in direction of motion. Between two successive changes in direction of motion, the particles continue to
move in the same direction for some variable time intervals. We have measured these time
intervals $\tau$ up to a fixed time $t$ or until the particles are annihilated (whichever is earlier) for every individual particle. 
Considering each particle of every configuration we have calculated the frequency $D(\tau)$.
Normalisation is done by adding  $D(\tau)$ for all $\tau$ and 
dividing it by the sum (probability of all $\tau$ should add up to 1).

\begin{figure}[h]
\includegraphics[width= 8cm]{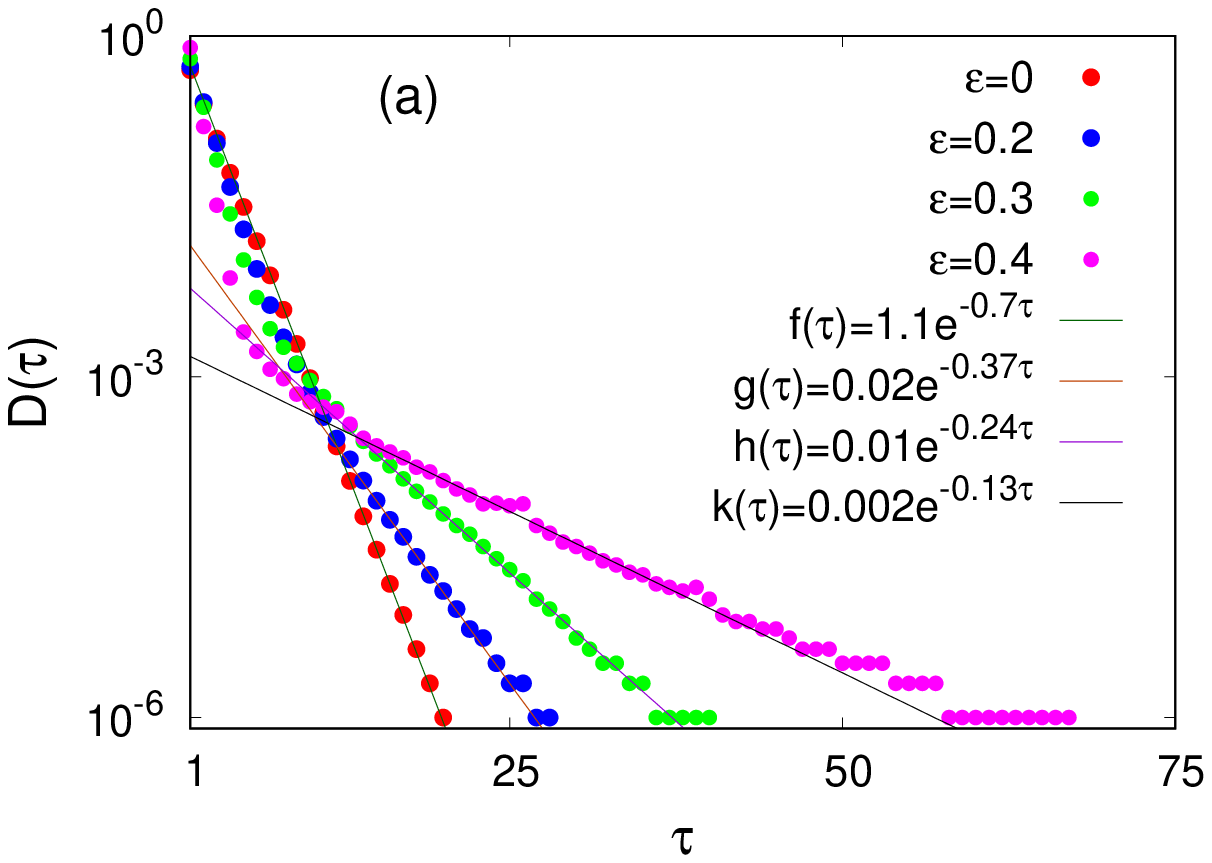}
\includegraphics[width= 8cm]{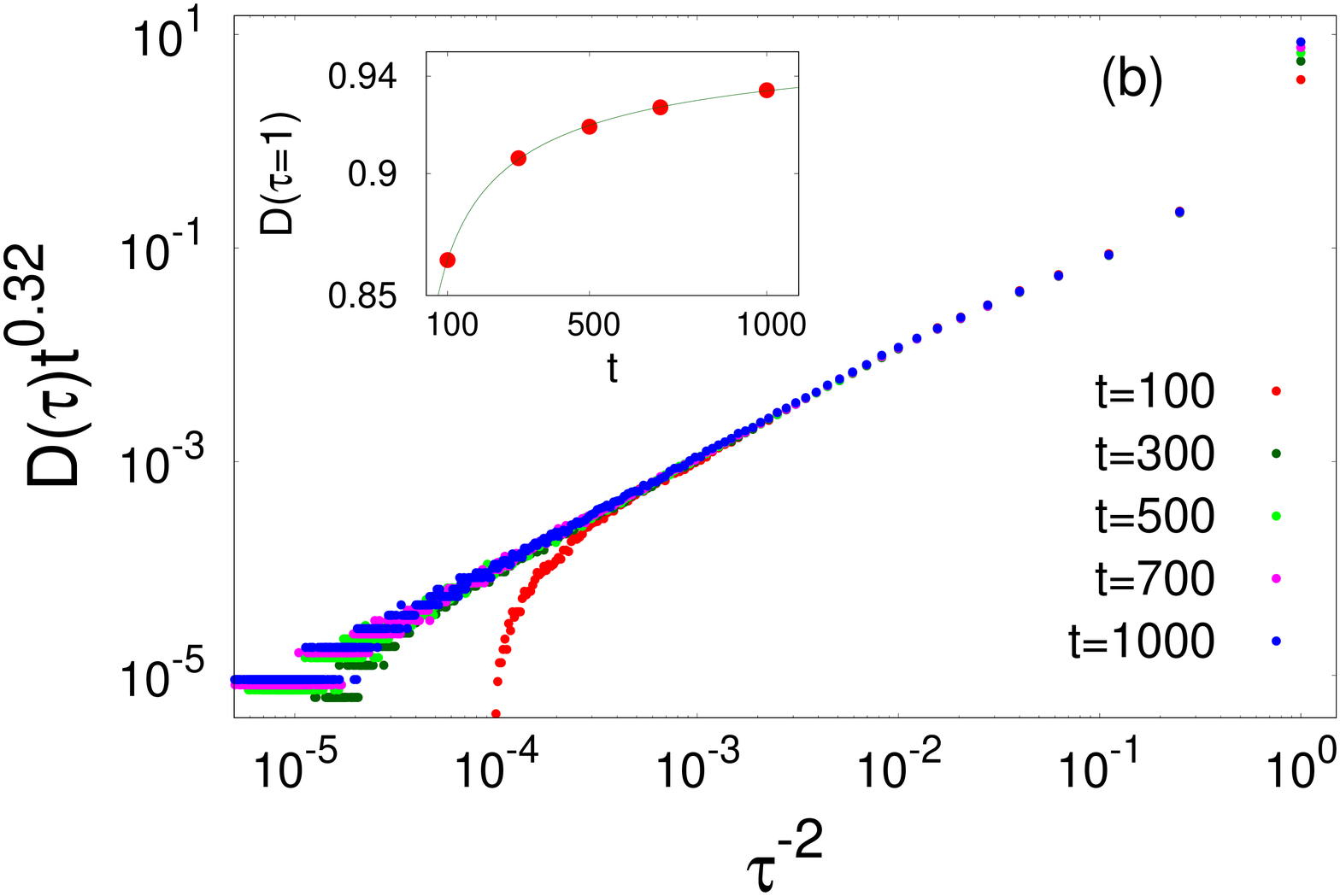}
\caption{(a): Variation of $D(\tau)$ over $\tau$ is shown in 
log-linear plot for several $\epsilon$.
The best fit lines are shown along with for different $\epsilon$ in the
same order. (b) shows the
data collapse of $D(\tau)t^{0.32}$ against $\tau^{-2}$ in 
log-log plot for $\epsilon=0.5$ for different $t$. 
Inset shows the values
of $D(\tau)$ at $\tau=1$ for different times for $\epsilon=0.5$ and are fitted to the form: $D(\tau=1,t) = 0.95[1-exp(-0.9t^{0.2})]$. These data are for a system size $L=12000$ averaged over 500 initial configurations.}
\label{timedistpara}
\end{figure}

For purely diffusive
motion ($\epsilon=0$), the probability of direction change at any time $t$
is $0.5$. So, the probability that in the time interval $\tau$,
the particles will not change their direction is given by the following equation:
\begin{eqnarray}
D(\tau)={0.5}^{2}({1-0.5})^{\tau},
\label{timeeq}
\end{eqnarray}
which reduces to an exponential form $D(\tau)\propto \exp[-\tau \ln 2 ]$ as shown in Fig. \ref{timedistpara}a.
For $0.5> \epsilon>0$, the tail of $D(\tau)$ shows an exponential decay; $D(\tau)\sim a'\exp(-b'\tau)$ (see Fig. \ref{timedistpara}a).
For $\epsilon=0.5$, no such exponential tail is observed, $D(\tau)$ instead shows a power law decay with $\tau$
with an exponent $2$, shown in Fig. \ref{timedistpara}b.

For $\epsilon=0.5$, at early times, we note that there are two kinds of motion,
some particles follow long trajectories in a straight line before getting 
annihilated or forming a dimer and other particles which quickly form  a dimer. At later times, only dimers remain (see Fig. 
\ref{snappara}b).  Hence  the  contribution to 
 $D(\tau)$ for  large values  of $\tau$ will come from the early times, i.e., the annihilation dominated regime. 
On the
other hand, at large times, dimer formation plays the key role 
when the particles typically change direction at every time contributing heavily to $D(\tau =1)$. 
Consequently we find $D(1)$ to grow in time  as shown in 
the inset of Fig. \ref{timedistpara}b. 

To explain the $\tau^{-2}$ dependence of the
tail, one can assume  $S(t) \approx S(t)_{ann}$, the contribution due to the annihilation only and use it 
to compute $D(\tau)$. 
Here, it may be mentioned that $D(\tau)$ for the  asynchronous case \cite{roy2020} also 
showed a $\tau^{-2}$ tail, where $S(t)$ was found to scale as $t^{-1}$.
Note that $S(t)_{ann}$ shows a leading order dependence as $1/t$ also in the parallel case (Eq. \ref{Stann}).
Hence one can derive the   
power law form  
of  $D(\tau) \propto \tau^{-2}$ for large $\tau$ in the same manner it was done in \cite{roy2020}. 

We also note that $D(\tau)$ has a distinct dependence on the particular time $t$ at which  it is calculated. 
$D(\tau=1)$ grows in time  and consequently  the magnitude of $D(\tau)$ for large $\tau$ 
decreases  (Fig. \ref{timedistpara}b).  In fact one can obtain a data collapse for the data at different times $t$ 
by plotting $D(\tau)t^{0.32 \pm 0.02}$ against $\tau $ such that  the behaviour of $D(\tau, t)$ for $\epsi = 0.5$ can be summarized as
\begin{equation}
D(\tau,t) = D(\tau=1,t)  \delta_{1,\tau} + {\rm const} ~ \frac{\tau^{-2}}{t^{0.32}}(1-\delta_{1,\tau}), 
\label{Dtaut}
\end{equation}
where we have fitted the growth of $D(\tau=1,t)$ by the function $D(1,t) = 0.95[1-\exp(-0.9t^{0.2})]$ shown in the inset of Fig. \ref{timedistpara}b. 

\section{Results for $\epsilon<0$}

\begin{figure}[h]
\centering
\includegraphics[width= 7cm]{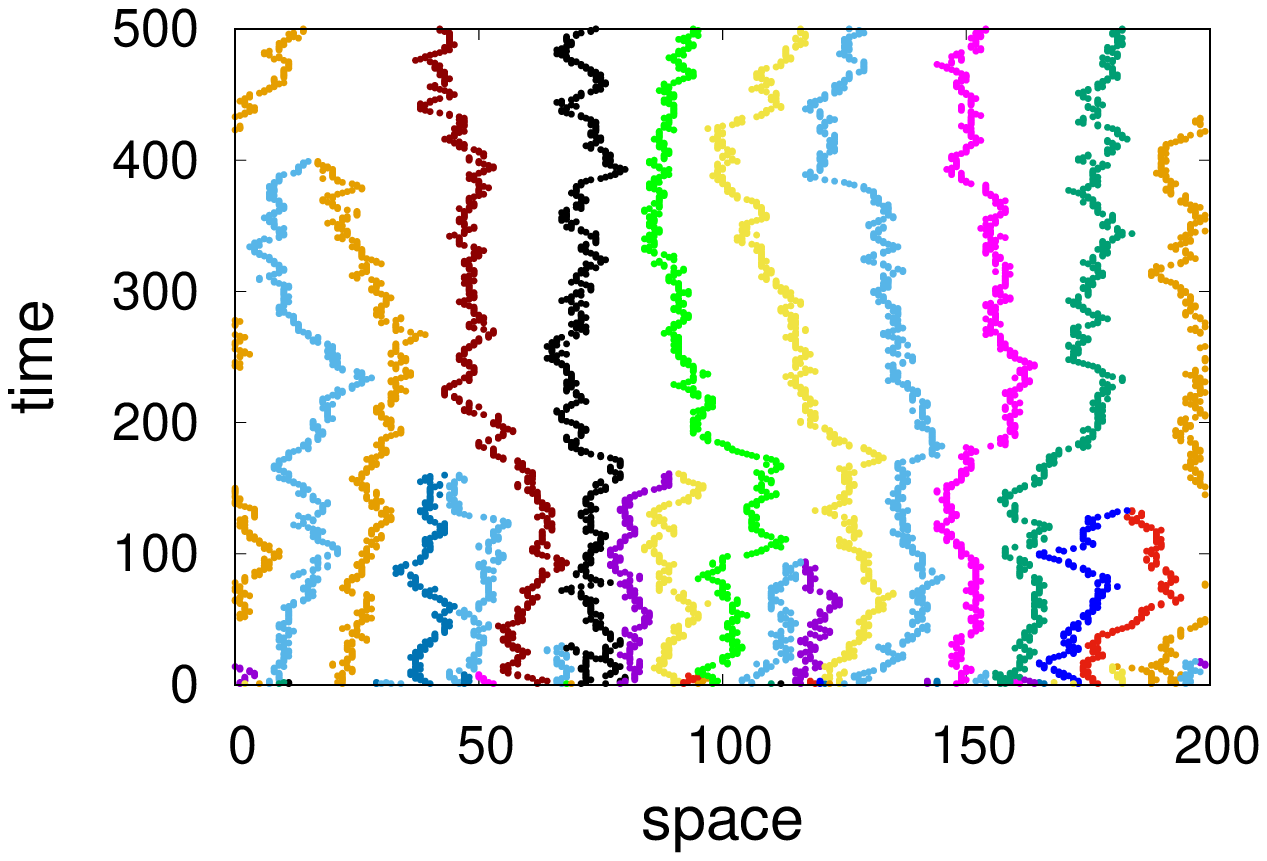}\hspace{0.3cm}
\includegraphics[width= 7cm]{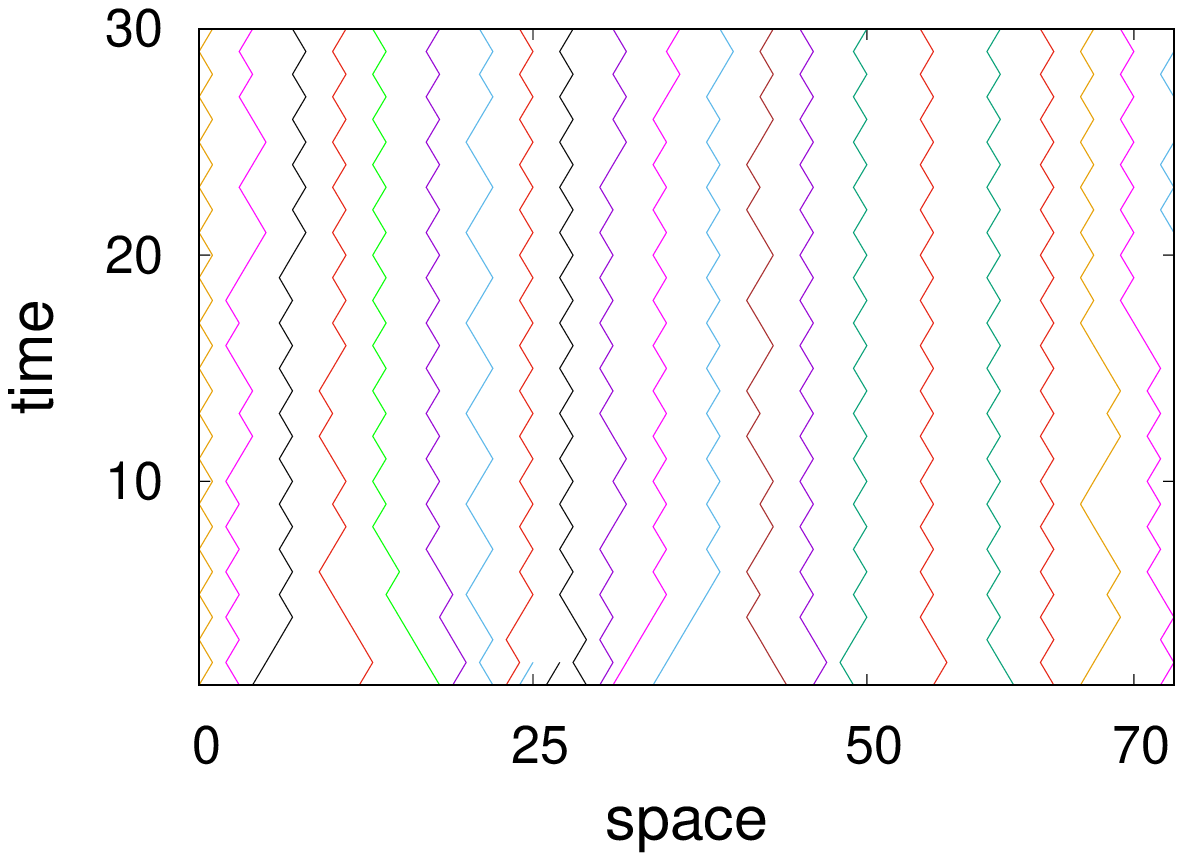}
\caption{Snapshots of the system at different times for $\epsilon=-0.1$
(left) and $\epsilon=-0.5$ (right). The
trajectories of different particles are represented by
different colors.}
\label{snapparaepneg}
\end{figure}

The particles have a bias $\epsilon$ to move towards their nearest neighbour.
As $\epsilon$ becomes negative, the particles tend to move away
from their nearest neighbour. Fig. \ref{snapparaepneg} shows the snapshots of the system
at different times for $\epsilon=-0.1$ and $-0.5$.  For $\epsilon=-0.5$, a particle always moves
away from its nearest neighbour, annihilation is extremely rare as it performs
 a nearly perfect  oscillatory motion at later times.
In general, since the particles are repulsive, even if two particles come close to form an isolated pair, they will move away from each other soon
such that no long lived dimer can exist here. 

\subsection{Bulk Properties}

\subsubsection{Fraction of surviving particle $\rho(t)$}

\begin{figure}[h]
\centering
\includegraphics[width= 9cm]{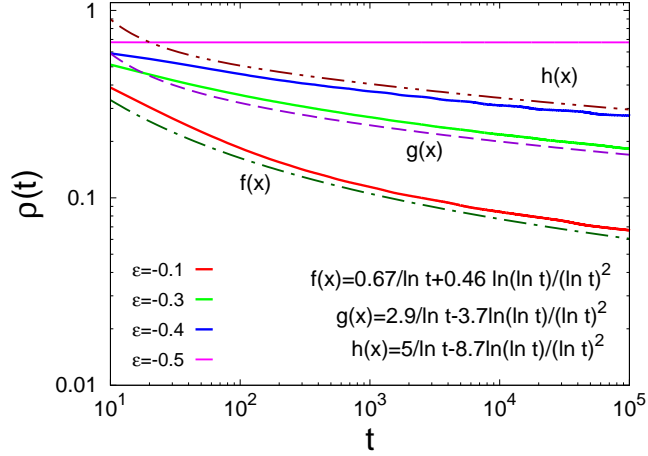}
\caption{Variation of fraction of surviving particles $\rho(t)$ in time $t$ for a system size $L=10000$
averaged over 200 initial configurations. Data are fitted to the form
of eq. (\ref{aliveeqneg}) with $\alpha,\beta$ values mentioned in the key. The best fitted lines (shifted slightly along $y$ axis for clarity) are shown along with for several $\epsilon$ in the same order .}
\label{aliveparaepneg1}
\end{figure}

For negative $\epsilon$, as the number of annihilation is smaller because of the
repulsion, the fraction  of surviving particles $\rho(t)$ shows a very slow decay in time that can be fitted to: 
\begin{equation}  
\rho(t)=\frac{\alpha}{\ln t}+\frac{\beta\ln (\ln t)}{(\ln t)^2},
\label{aliveeqneg}
\end{equation} 
where $\alpha$, $\beta$ are $\epsilon$ dependent. 
Fig. \ref{aliveparaepneg1} shows 
the data for $\rho(t)$ against $t$ for several $\epsilon$. 
Here it must be mentioned that $\epsi = -0.5$ is a special point for which  eq. \ref{aliveeqneg} is not valid. 
For $\epsi= -0.5$,  the particles achieve a equidistant configuration at large time and every particle performs a to and fro movement (as the dynamical rule ensures that each particle has to undergo a displacement); no annihilation will take place and  $\rho(t)$ rapidly saturates to a constant value ${\mathcal {O}}(10^{-1})$. 

\subsubsection{Persistence probability $P(t)$}

The persistence probability $P(t)$ shows an interesting behaviour for $\epsi < 0$. 


For any $\epsi \neq -0.5$, it shows a fast decay with time, however,   
the magnitude of the 
persistence probability shows a non-monotonic behaviour. For $0 > \epsi >  -0.4$, it decreases as $\epsilon$ decreases, 
but as $\epsilon$ becomes more negative,
the decay rate becomes slower.
$P(t)$ shows a stretched exponential decay in time and the data can 
be fit to the following form
\begin{equation}
P(t)=q_0\exp(-qt^r).
\label{pereqneg}
\end{equation}

\begin{figure}[h]
\centering
\includegraphics[width= 9cm]{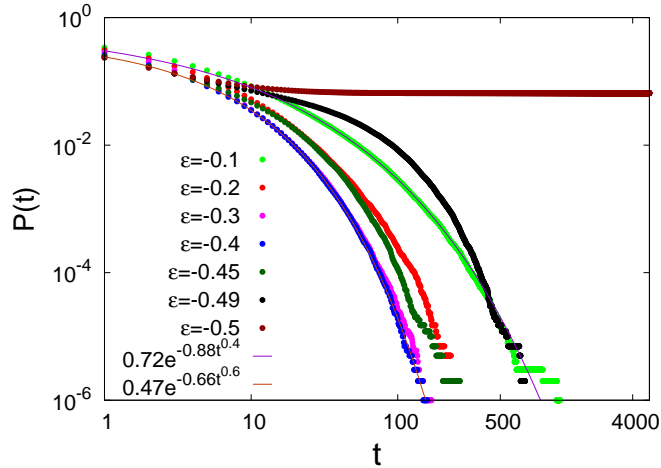}
\caption{Variation of persistence probability $P(t)$ with $t$ for different $\epsilon$. The
best fit lines are shown for $\epsilon=-0.1$ and $\epsilon=-0.4$. These data are for a system size $L=8000$
averaged over 200 realisations.}
\label{persisparaepneg}
\end{figure}

For $\epsilon=-0.5$, the  movement of the particles is restricted as they perform nearly oscillatory motion,
as shown in Fig. \ref{snapparaepneg},
most of the sites remain unvisited. Therefore, $P(t)$ shows a very slow decay at the initial few steps and then becomes a constant in time as
shown in Fig. \ref{persisparaepneg}. 

The  results for  the     persistence probability shows that  for $|\epsi| < \sim 0.4$, the annihilation decreases such that  more number of particles remain in the system which  display a certain degree of mobility, 
thereby decreasing the persistence probability.  However,   for   $|\epsi| > \sim 0.4$, the particle 
mobility gets seriously restricted, such that, although a larger number of 
particles survive in the system,    more sites remain unvisited. 
This indicates an interesting crossover behaviour in the motion of the 
particles as $|\epsi|$  increases,  captured by the behaviour of $P(t)$. 

$P(t)$ shows a similar stretched exponential decay in case of asynchronous dynamics for $\epsilon<0$. However,
$P(t)$ decreases monotonically as $\epsilon$ becomes more negative in the asynchronous case.

\subsection{Tagged particle properties}

\subsubsection {Probability distribution $\Pi(x, t)$}

Probability distribution $\Pi(x, t)$ retains its Gaussian form when $\epsilon$
is negative. But the scaling variable  $x/t^\nu$ is accompanied by a non-unique value of $\nu$ that decreases  from 0.5
monotonically as $\epsilon$ becomes more negative.

\begin{figure}[h]
\centering
\includegraphics[width=11cm]{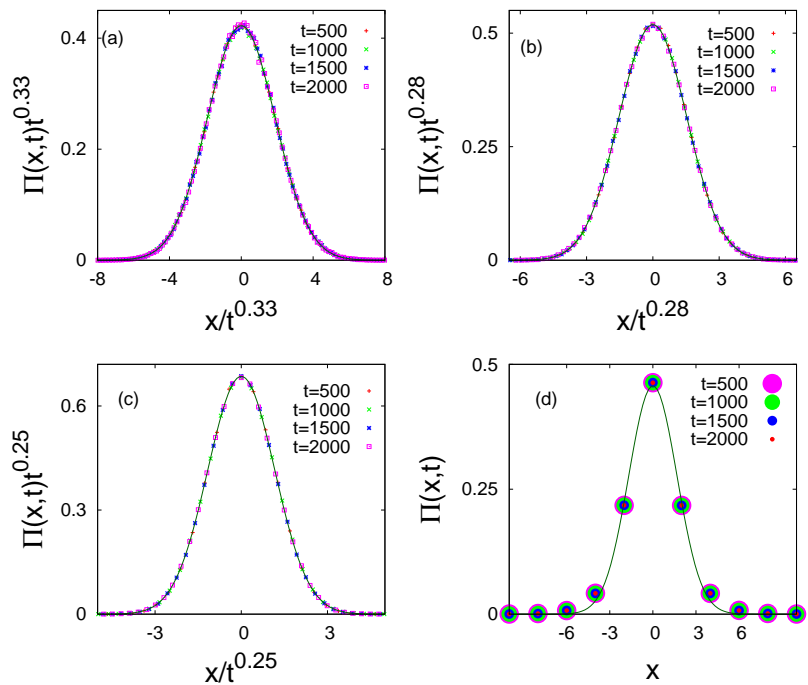}
\caption{Data collapse of $\Pi(x,t)$ is shown for $\epsilon=-0.1$ (a), $\epsilon=-0.3$ (b), $\epsilon=-0.45$ (c) and $\epsilon=-0.5$ (d).
Data are fitted to the Gaussian distribution form. Scaling functions are $f(x/t^{0.33})= 0.42\exp[-0.14(x/t^{0.33})^2]$, 
$g(x/t^{0.28})=0.52\exp[-0.21(x/t^{0.28})^2]$,
$k(x/t^{0.25})=0.69\exp[-0.37(x/t^{0.25})^2]$,  $h(x)=0.18\exp(-0.46 x^2)$ for (a), (b), (c) and (d) respectively,
shown in the figure. These data are for system size $L=12000$ and the number of configuration studied was 500.}
\label{distriparaepneg}
\end{figure}

Fig. \ref{distriparaepneg} shows collapsed data at different times when
$\Pi(x,t)t^\nu$ is plotted against $x/t^\nu$.    
For $\epsilon=-0.5$, the particles attain a equidistant configuration; but according to the dynamical
rule, as the particles must make a move, they only perform a back and forth movement (see Fig. \ref{snapparaepneg}).
As a result,  the probability distribution 
$\Pi(x,t)$ becomes time independent after a brief transient, shown in Fig. 
\ref{distriparaepneg}(d).

\begin{figure}[h]
\centering
\includegraphics[width=8cm]{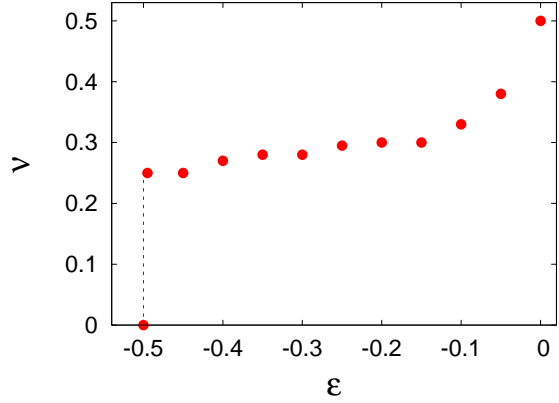}
\caption{Variation of scaling variable $\nu$ with $\epsilon$.}
\label{alphafig}
\end{figure}

Fig. \ref{alphafig} shows the value of 
$\nu$ against $\epsilon$ that decreases  from 0.5
monotonically as $\epsilon$ becomes more negative. At $\epsi = -0.5$, there is 
a sharp discontinuity in its value as it falls to zero  from a value 
$\sim 0.25$. 

\subsubsection {Probability of direction change $S(t)$}

\begin{figure}[h]
\includegraphics[width= 8cm]{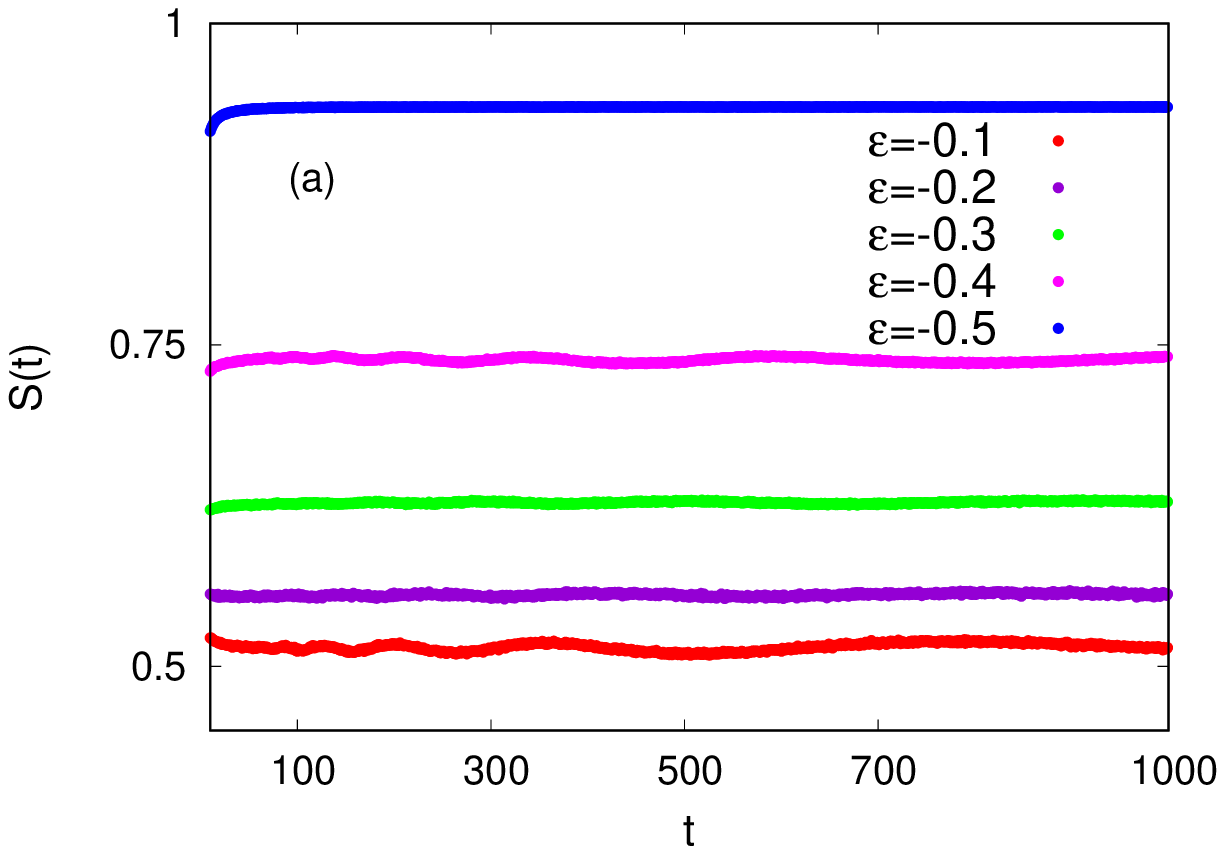}
\includegraphics[width= 8cm]{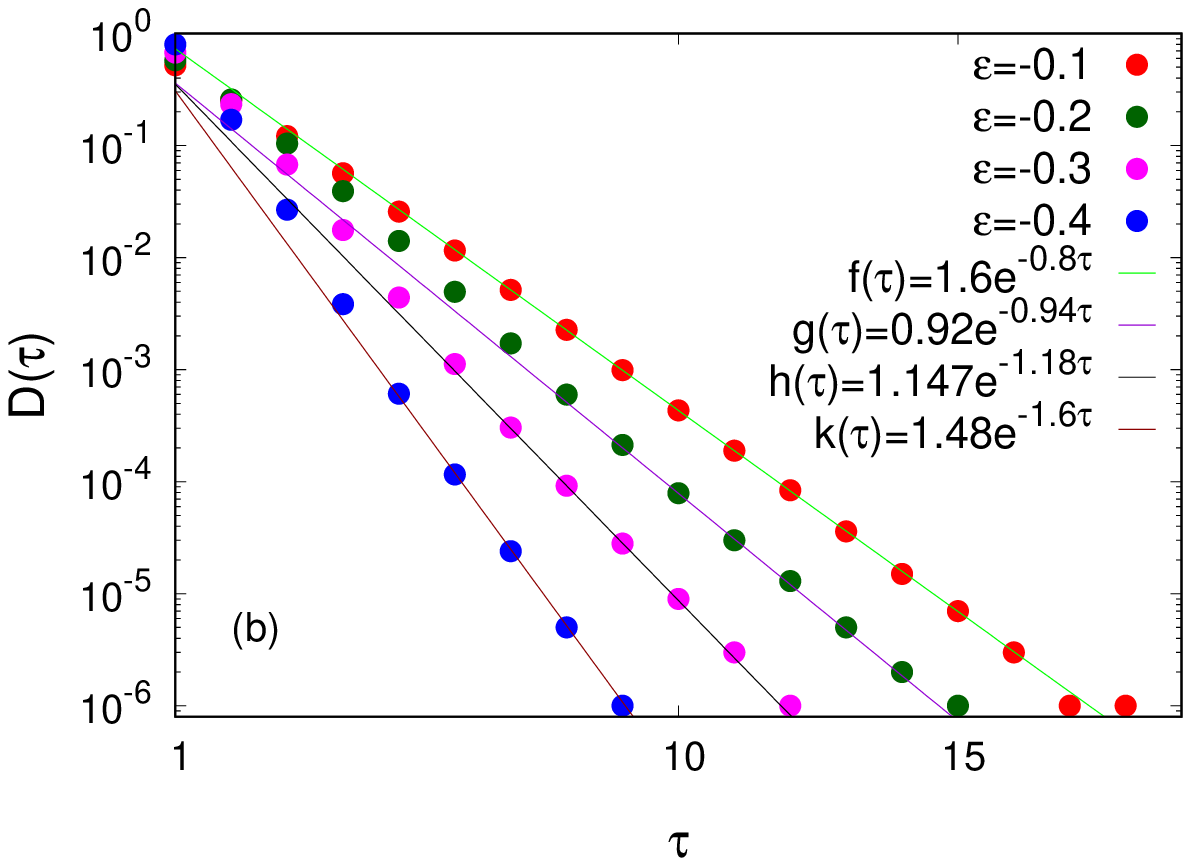}
\caption{(a) shows the probability of direction change $S(t)$ of a tagged particle at time $t$ for
different $\epsilon$. (b) shows Variation of $D(\tau)$ over $\tau$ for different $\epsilon$ in log-linear plot. 
The best fit lines are shown along with for different $\epsilon$ in the
same order. These data are for a system of size $L=10000$ taking average over 500 realisations.}
\label{dirchangeparaepneg}
\end{figure}

For $\epsilon<0$, $S(t)$ attains almost a constant value that increases systematically with the magnitude of  
$\epsi$,  shown in Fig. 
\ref{dirchangeparaepneg}(a).
Here, as the annihilation factor is less relevant, especially at later times, the  change in direction of motion occurs due to the 
repulsion between the neighbouring particles mainly in the following manner:
as $\epsilon$ decreases, the repulsive factor becomes stronger
and the particles tend to avoid  their nearest neighbours. A change in the
direction  can occur  if  the other neighbour comes closer as a result.
At the extreme limit $\epsilon=-0.5$, this happens at every step such that the change in direction is maximum.

\subsubsection{Distribution of time interval spent without change in direction of motion $D(\tau)$}

For $\epsilon<0$, as $S(t)$ becomes constant,
$D(\tau)$ is expected to show an exponential decay. 
$D(\tau)$ shows a faster decay with  $\tau$ as $\epsi$ becomes more negative. 
 For $\epsilon=-0.5$,
since  the particles change their direction of motion more often, 
 $D(\tau)$   decays in the fastest manner. 

The tail of the distribution $D(\tau)$ can be fit to the form 
\begin{equation}
D(\tau)=c\exp(-d\tau),
\label{timeparanegeq}
\end{equation}
and the results  are shown in Fig. \ref{dirchangeparaepneg}(b).

\section{Comparison of parallel and asynchronous update}
\label{comparison}

Having obtained  all the results for the entire range of $\epsilon$, one can now make a comparison between the results for the  parallel and asynchronous dynamics for the $A + A \to \emptyset$ model where the particles have a bias to move toward their nearest neighbours. The comparison of the different properties are 
presented  in Table \ref{table}. 
We note that while for $\epsi > 0$, the results are significantly different, the negative $\epsi$  results are 
almost independent of the particular update used.

For $\epsi > 0$, both $\rho(t)$ and $P(t)$ are modulated by a factor of $\ln t$. This is attributed to the  
dimerisation that is present only for  the parallel updating case in this particular model and for $\epsi > 0$.

Another  notable  difference is in the behaviour of the 
probability distribution,  and 
this is a suitable juncture to further analyse the behaviour of $\Pi(x,t)$ with the parallel updating scheme.  
For the asynchronous update we obtained a double peaked structure which was ascribed to the dominantly ballistic walkers existing 
in the later time regime. Here instead, we get a single peaked structure (see Fig. \ref{distpara}). To understand this, we first  
consider the extreme case of $\epsi = 0.5$. The snapshots of Fig. \ref{snappara} show that 
 a considerable  fraction of  the particles quickly form  dimers while some 
particles    follow a ballistic path, in either direction.  
The particles which form dimers remain close to their origins and thus contribute to $x \approx 0$ giving rise to the peak at $x=0$.
The ballistic particles will contribute towards $|x| \gg  0$.
We conjecture that the heavy tail of the  distribution  $\Pi(x,t)$ is 
connected to these ballistic particles. 

That the ballistic walkers remain in the system is  corroborated by the fact that $D(\tau)$ shows a power law behaviour for large $\tau$ for $\epsi = 0.5$, shown in Fig. \ref{timedistpara}b. 
For smaller $\epsi$, dimers are not formed easily 
and the single peaked behaviour is due to the 
enhanced probability of direction change $S(t)$ which results in much 
smaller net displacements.   
On the other hand, in the  asynchronous case,  the direction change is much less  probable, the particles 
perform an overall ballistic walk even for small $\epsi$ and hence the double peaked structure is present for all $\epsilon$. 
The ballistic walk occurs maximally for $\epsi = 0.5$. So  the exponent 
associated with the power law behaviour of the scaled $\Pi(x,t)$ is least for $\epsi = 0.5$ and in general decreases as $\epsi$ increases. 
The  width of the power law region 
increases with $\epsi$ due to  the same reason.

For $\epsi < 0$, there are some differences in the persistence
probability and the exponent $\nu$ occurring in  the scaling variable of the probability distribution $\Pi$, with respect to their variations with  $\epsilon$. 
The probability distribution $\Pi(x,t)$ is Gaussian for $\epsilon<0$ independent of the dynamics used. For $\epsilon=-0.5$,
the scaling factor $\nu=0$ for the parallel dynamics (reported in the present work) and $\nu=0.25$  for asynchronous dynamics as expected for  repulsive
random walkers \cite{arratia}.
 However, for $\epsilon$ very close to $-0.5$, $\nu$ shows a value 0.25 for parallel dynamics also; only at $\epsilon=-0.5$,
$\nu$ shows a discontinuity as $\Pi(x,t)$ becomes time independent. 
Clearly this is because  
the parallel updating scheme leads to  oscillatory motions as $\epsi \to -0.5$, in the  asynchronous update, there is no
such oscillation. In that sense, the motion of the particles are more correlated for the parallel dynamics.
A discontinuity is also noted in the behaviour of  $\rho(t)$ for $\epsi < 0$; at $\epsi = -0.5$, one cannot fit it to the 
form eq. \ref{aliveeqneg}.  

\section {Concluding remarks}

 In this paper, we have studied the effect of the synchronous
(parallel) dynamical rule on the $A+A \to \emptyset$ model in one dimension, where
the particles move towards their nearest neighbour, to check how far the parallel dynamics
change the results.
The probability to move towards the nearest neighbour is taken
parametrically as $0.5 + \epsilon$ where $-0.5 \leq \epsilon \leq 0.5 $. 

The properties of the model have been summarized in Table \ref{table}. 
For $\epsi > 0$  the results depend 
strongly on the dynamical rule used; synchronous or asynchronous.   It is the presence of long surviving 
dimers, composed of particles making a flip-flop motion 
due to  the parallel dynamical rule,  that mostly  gives rise to 
a number of interesting variations in the relevant quantities. 
A $\ln t$ factor is seen to modulate the power law decay of the 
particle density and persistence probability when compared to the results of the 
asynchronous update. This is   attributed to the the presence of the dimers. In order to 
confirm this, simulations with an initial condition with no randomness in initial position of the particles was considered which 
does not allow dimers.  Here, particles occupy either odd or even sites. 
In fact this case simply coincides with the asynchronous dynamics as  the particles 
cannot cross each other and is therefore 
not a surprise.  Thus  it appears that   dimers could be the key factor  responsible for altering the scaling behaviour for the random initial condition. Even if dimers are not permanent for $\epsi < 0.5$, they are long lived enough to affect the dynamics in the scaling regime.
 In this context it may be added that dimer formation is possible in principle with 
other kinds of stochastic walks   and even with asynchronous 
dynamics, e.g., when step lengths $> 1$ is allowed.  It will be an interesting issue to see whether the scaling behaviour is affected similarly by their presence in these models.

 As discussed in the introduction, the results depend on the odd/even- ness
of the lattice and the number of particles as well as on the initial
condition. Our results are applicable for a random initial condition with even number of 
particles to begin with and a lattice size which is a multiple of 4. It may also 
be added that the initial condition of particles sitting at only odd/even sites can only be possible as long as the initial density is less than or equal to 1/2. 

In addition to the bulk properties, we have  analysed how the tagged particle properties like $S(t)$ and $D(\tau, t)$ are dependent on the presence of dimers  for $\epsi = 0.5$.  The results 
 reveal 
the crossing over of the system from annihilation dominated to dimer dominated regimes. In this context, let us recall that a  crossover 
from a annihilation to diffusion dominated regime for the asynchronous case  was found recently \cite{roy2020}.

Another intriguing result is that we find that the persistence exponent in the parallel case seems to
be twice of the one found in the asynchronous case for $\epsi = 0$. 
Such doubling of persistence exponent could be proved for the Ising Glauber or Potts model  with parallel dynamics. Although for asynchronous dynamics, the
$A + A \to \emptyset$ model with  $\epsi = 0$ and   the Ising Glauber model are identical, with parallel dynamics, such a correspondence no longer exists. 
So the  result obtained here for the persistence exponent at $\epsi = 0$ 
 for the parallel dynamics
is  neither naively expected nor simply obvious.

It is  understandable why for negative $\epsi$, the  results for the dynamical quantities are independent of the updating scheme apart from
subtle differences in their $\epsi$ dependence. 
The choice of the dynamical scheme
affects the annihilation process significantly.  For $\epsi < 0$, as the particles repel each other, they hardly come into contact to
annihilate each other and hence the results are more or less similar.

Acknowledgement: The authors thank DST-SERB project, File no. EMR/2016/005429 
(Government of India) for financial support. Discussion with Soham Biswas is also acknowledged.

\begin{small}
\begin {table}[h]
\caption{\textbf {Leading order time and $\epsilon$ dependence of several quantities in one dimensional $A+A \to \emptyset$ model.} Results for asynchronous dynamics are quoted from references \cite{soham_ray_ps2011,roy2020,roy_ps2020,park2}. Notation used: generic.}
\begin{center}
\begin{tabular}{ |c|c|c| }
\hline
	    	&						&							\\
  	    	& 		Asynchronous 			& 	Parallel  					\\
		&						&							\\ 
\hline
 		&  $t^{-0.5}$ for $\epsilon=0$   		&  	$ t^{-0.5}$ for $\epsilon=0$			\\
$\rho(t)$  	&   $ t^{-1}$ for  $\epsi >0$ 			& 	$ \ln t/t$  for  $\epsi >0$		\\
  		&  						& 	$t < t^*$		\\
		& Leading order term $\frac{\alpha}{\ln t}$ for $0> \epsi \neq -0.5 ^\dagger$ 	& 	Leading order dependence $\frac{\alpha}{\ln t}$ for $0 > \epsi \neq -0.5 $			\\
 		&						& 	saturates rapidly for $\epsilon=-0.5$		\\
\hline
		&						&							\\
		&  $ t^{-0.375}$ for $\epsilon=0$   		&  	$  t^{-0.75}$ for $\epsilon=0$	 		 \\
 $P(t)$ 	&  $ t^{-0.235}$ for $\epsilon>0$   		&  	$ t^{-0.72}\ln t$ for $\epsilon>0$			\\
 
 		&   $ a\exp(-bt^c)$ for $\epsilon<0 $ 		&      $ a\exp(-bt^c)$ for $\epsi  < 0 \neq  -0.5$	\\
 		&						& 	saturates rapidly for $\epsilon=-0.5$		\\
 		&						& 	 A crossover behaviour noted 		\\
\hline

		&						&							\\
		&  Scaling factor  $x/t^\nu$   in all cases	&  Scaling factor  $x/t^\nu$ in all cases		\\
		
		& $\Pi$ Gaussian for $\epsi = 0$ and $\epsilon < 0$   & $\Pi$     Gaussian for $\epsi = 0$ and $\epsi < 0$  	\\

$\Pi(x,t)$ 	& $\nu=0.5$ for $\epsilon=0$ 		&          $\nu=0.5$ for $\epsilon=0$ 		\\
		& $\nu$ decreases with $\epsi$ for $\epsi<0$ &       $\nu$ decreases with $\epsi$ for $\epsi<0$    \\
		&						&	with a discontinuity at $\epsi = -0.5$	         \\ 
		&						&							\\
		& $\Pi$ double peaked  for $\epsilon>0$ 		&   $\Pi$  Non Gaussian single peaked for $\epsilon>0$        \\
		& $\nu=1$ for $\epsi>0$ 			&       $\nu=0.55 \pm 0.05$ for $\epsi>0$ 			 \\

		&						& $\Pi(x,t)t^{\nu}$ shows a power law regime for \\
                &						&   				large values of $x/t^{\nu}$ \\
\hline

		&						&							\\
		&   $const$ for $\epsi = 0$ and $\epsi < 0$   	& 	$const$   for $\epsi = 0$ and $\epsi < 0$ 	\\
$S(t)$ 		&     $t^{-1}$ up to $t^*$ ($\epsi > 0$) 	&      constant at very large times 			\\ 		 
		&	$t^* \propto 1/(0.5-\epsi)$		&	Non monotonic behaviour for $\epsi$ close to 0.5 \\					 &  $S(t)$ decreases as $\epsilon$ increases 	& 	$S(t)$ increases as $\epsilon$ increases	\\
\hline
		&						&							\\
		& $\exp(-a\tau)$ for $\epsi = 0$ and $\epsi<0$  &	$\exp(-a\tau)$ for $\epsi = 0$ and $\epsi < 0$  \\
$D(\tau)$ 	& $\tau^{-2}$ upto $\tau^*$ ($\epsilon>0)$ 	& $\exp(-a\tau)$ for large $\tau$ for $\epsi>0 \neq 0.5$\\
 		& $\tau^* \propto 1/(0.5-\epsi)$ 		&  		$\tau^{-2}$ for $\epsi=0.5$ 		\\
		&						&							\\
\hline
\hline

\end{tabular}
\label{table}
\end{center}
$^{\dagger}$  This result is quoted from \cite{park2}. In \cite{roy_ps2020}, the numerical results 
were shown to fit a power law form. 
\end {table}
\end{small}

\end{document}